\documentclass[11pt]{article}

\pdfoutput=1

\usepackage{stackrel}
\usepackage{mathrsfs}
\usepackage{fancyref}
\usepackage{graphicx}
\usepackage{enumerate}
\usepackage{mdwlist}
\usepackage{caption}
\usepackage{fancybox}
\usepackage{pifont}
\usepackage[all]{xy}
\usepackage{makecell}
\usepackage{epigraph}
\usepackage{pifont}
\usepackage{dsfont}
\usepackage{relsize}
\usepackage{enumitem}
\usepackage{float}
\usepackage{pifont}

\usepackage{epsfig}
\usepackage{graphicx}

\usepackage{latexsym,times}
\usepackage{amssymb,amsmath}
\usepackage{amsxtra}
\usepackage{amscd}
\usepackage{amsthm}
\usepackage{amsfonts}
\usepackage{dsfont}
\usepackage{xcolor}
\usepackage{cancel}
\usepackage{framed}

\usepackage{esint}

\usepackage{comment}

\usepackage{enumitem}
\usepackage{float}

\usepackage{hyperref}
\usepackage[square,sort&compress,comma,numbers]{natbib}
\makeatletter

\@addtoreset{equation}{section} \makeatother

\newcommand\mr{\mathring}
\newcommand\Hf{\mathbb{H}}

\newcommand{\deltaM}{\delta_{\scriptscriptstyle{\rm{M}}}}
\newcommand{\deltaw}{\delta_{\scriptscriptstyle{\rm{\bf GL}}}}

\newcommand{\half}{{{\textstyle\frac{1}{2}}}}
\newcommand{\quarter}{{{\textstyle\frac{1}{4}}}}
\newcommand{\be}{\begin{equation}}
\newcommand{\ee}{\end{equation} }
\newcommand{\beqa}{\begin{eqnarray} }
\newcommand{\eeqa}{\end{eqnarray} }
\newcommand{\ba}{\begin{array}}
\newcommand{\ea}{\end{array}}
\newcommand{\bpm}{\begin{pmatrix}}
\newcommand{\epm}{\end{pmatrix}}

\newcommand{\Spin}{\mathbf{Spin}}
\newcommand{\GL}{\mathbf{GL}}

\newcommand{\rmd}{{\rm d}}

\newcommand{\rmD}{{\rm D}}

\newcommand{\ODD}{\mathbf{O}(D,D)}

\newcommand{\brfD}{\bar{\frak{D}}}
\newcommand{\hfD}{\widehat{\frak{D}}}

\newcommand{\hOmega}{\widehat{\Omega}}

\newcommand\cA{{\cal A}}
\newcommand\cB{{\cal B}}

\newcommand\cH{{\cal H}}

\newcommand\cJ{{\cal J}}

\newcommand\cL{{\cal L}}
\newcommand\cM{{\cal M}}

\newcommand\cP{{\cal P}}

\newcommand\cX{{\cal X}}

\newcommand\brcP{{\bar{\cal{P}}}}

\newcommand\hcL{{\hat{\cal L}}}

\newcommand\fD{{\mathfrak{D}}}

\newcommand\hGamma{\hat{\Gamma}}

\newcommand\hE{\widehat{E}}

\newcommand\hS{{\hat{S}}}

\newcommand\dis{\displaystyle}

\def\tx{\tilde{x}}

\def\tpartial{\tilde{\partial}}

\def\omegaT{\omega_{{\scriptscriptstyle{T}}}}

\def\brc{\bar{c}}

\def\bri{\bar{\imath}}
\def\brj{\bar{\jmath}}
\def\brk{{\bar{k}}}

\def\brn{{\bar{n}}}
\def\brp{{\bar{p}}}

\def\brw{{\bar{w}}}

\def\brUpsilon{{{\bar{\Upsilon}}}}

\def\brV{{\bar{V}}}

\def\brX{{\bar{X}}}
\def\brY{{\bar{Y}}}
\def\brP{{\bar{P}}}

\def\hPhi{{\widehat{\Phi}}}

\newcommand{\na}{{\nabla}}
\newcommand{\trd}{{\bigtriangledown}}

\def\MD{\mathds{D}}
\def\hMD{\widehat{\mathds{D}}}

\newcommand\So{S_{\scriptscriptstyle{{(0)}}}}

\newcommand\mcB{\mathcal{B}}

\newcommand\twist{\mr}

\def\mT{\twist{T}}
\def\mP{\twist{P}}
\def\mS{\twist{S}}
\def\mSo{\twist{S}_{\scriptscriptstyle{{(0)}}}}

\def\mV{\twist{V}}

\def\mbrP{\twist{\brP}}

\def\mcJ{\twist{\cJ}}
\def\mcH{\twist{\cH}}

\def\mcP{\twist{\cP}}

\def\mbrcP{\twist{\brcP}}

\def\mGamma{\twist{\Gamma}}

\newcommand\hH{\widehat{\Hf}}

\newcommand{\darkblue}[1]{\color[rgb]{0.2,0.0,0.8}{#1}}

\newcommand{\bea}{\begin{eqnarray}}
\newcommand{\eea}{\end{eqnarray}}

\definecolor{rougef}{rgb}{0.56,0,0}		
\definecolor{vertf}{rgb}{0,0.5,0}		
\definecolor{bleuf}{rgb}{0,0,0.8}

\begin{document}

\begin{titlepage}

\title{\vskip -100pt
\vskip 2cm 
{{{Remarks on the non-Riemannian sector  in  Double Field Theory }}}}

\author{\sc Kyoungho Cho\quad and\quad Jeong-Hyuck Park}
\date{}
\maketitle 
\begin{center}
Department of Physics, Sogang University, 35 Baekbeom-ro, Mapo-gu,  Seoul  04107, Korea\\
~\\
\texttt{khcho23@sogang.ac.kr \qquad park@sogang.ac.kr }\\
~\\
~\\

\end{center}
\begin{abstract}
\vskip0.1cm  
\centering\begin{minipage}{\dimexpr\paperwidth-7.4cm}
\noindent  Taking  $\mathbf{O}(D,D)$ covariant field variables as its   truly   fundamental  constituents,   Double Field Theory  can  accommodate  not only  conventional  supergravity but also  non-Riemannian gravities that may  be      classified  by two non-negative integers, $(n,\bar{n})$.  Such non-Riemannian   backgrounds    render a  propagating  string    chiral and anti-chiral over  $n$ and $\bar{n}$ dimensions respectively.    Examples include, but are not limited to,    Newton--Cartan, Carroll, or Gomis--Ooguri.  Here   we  analyze        the variational principle   with care  for a generic $(n,\bar{n})$  non-Riemannian sector.  We recognize     a  nontrivial    subtlety for ${n\bar{n}\neq 0}$ that  infinitesimal variations generically  include those which change  $(n,\bar{n})$.  This seems to suggest  that the various non-Riemannian gravities should  better be identified  as different solution sectors of Double Field Theory rather than  viewed  as independent theories.      
Separate verification of our results as  string   worldsheet   beta-functions   may enlarge the scope of the string    landscape far   beyond Riemann.

\end{minipage}
\end{abstract}

\thispagestyle{empty}

\end{titlepage}

{\small{\tableofcontents}}

\section{Introduction}
This paper is a  sequel to \cite{Morand:2017fnv} which proposed to classify all  the possible geometries of Double Field Theory (DFT)~\cite{Siegel:1993xq,Siegel:1993th,Hull:2009mi,Hull:2009zb,Hohm:2010jy,Hohm:2010pp} by  two non-negative integers, $(n,\brn)$.  The outcome  ---which we shall review in section~\ref{SECnbrn}---   is that only the case of $(0,0)$ corresponds to  conventional supergravity  based on Riemannian geometry. Other generic   cases of  ${(n,\brn)\neq(0,0)}$ do not  admit    any   invertible   Riemannian metric    and hence are \textit{non-Riemannian} by nature.  Strings propagating on these backgrounds  become chiral and anti-chiral over  $n$ and $\bar{n}$ dimensions respectively.

The non-Riemannian  property is  a point-wise or local statement~\cite{Lee:2013hma,Park:2014una,Ko:2015rha,Berman:2019izh,Sakatani:2019jgu} and differs from the    global notion of `non-geometry'~\cite{Obers:1998fb, Hull:2004in,  Hull:2005hk, DallAgata:2007egc, Blumenhagen:2011ph} which is also well described by DFT~\cite{Andriot:2011uh, Andriot:2012an, Blumenhagen:2012nt, Dibitetto:2012rk, Malek:2013sp, Hassler:2014sba, Cederwall:2014opa,Bakhmatov:2016kfn, Heller:2016abk, Lee:2016qwn, Chatzistavrakidis:2018ztm, Marotta:2018myj, Plauschinn:2018wbo, Deser:2018flj,Otsuki:2019owg}.  Possible examples of non-Riemannian geometries include  Newton--Cartan geometry~\cite{Cartan:1923zea,Kuenzle:1972zw,Duval:1984cj} as $(1,0)$, stringy Newton--Cartan~\cite{Andringa:2012uz} as $(1,1)$,  (wonderland)  Carroll geometry~\cite{Henneaux79,Duval:2014uoa} as $({D-1},0)$, and   non-relativistic Gomis--Ooguri  string theory~\cite{Gomis:2000bd} as $(1,1)$. These  are  of  continuous  interest, \textit{e.g.}~\cite{Milne:1934,Duval:1993pe,Christensen:2013lma,Bekaert:2014bwa,
Bekaert:2015xua,Bergshoeff:2016lwr,Hartong:2016yrf,Bergshoeff:2018yvt,
Hansen:2018ofj,Harmark:2018cdl,Bergshoeff:2018vfn,Morand:2018tke,
Gomis:2019zyu,Hansen:2019svu,Gallegos:2019icg,Harmark:2019upf,
Bergshoeff:2019pij,Blair:2019qwi,Pereniguez:2019eoq}.  Further, the fully $\ODD$ symmetric vacua of Double Field Theory turn out to be `maximally' non-Riemannian,  being of  either $(D,0)$ or $(0,D)$ type, compelling   string to be completely chiral or anti-chiral.  A remarkable  insight   from  \cite{Berman:2019izh}  is that, the ordinary  Riemannian spacetime arises after  spontaneous symmetry  breaking of these  fully $\ODD$ symmetric vacua  while identifying  the Riemannian metric, $g_{\mu\nu}$, as a Nambu--Goldstone boson.

 In this work   we attempt to explore   the dynamics of the generic $(n,\brn)$ sector in Double Field Theory.  We analyze with care the relevant variational principle  and  recognize  a   nontrivial  subtlety: when $n\brn\neq0$, the resulting Euler--Lagrangian equations of motion depend whether the variations of  the action keep the values of $(n,\brn)$ fixed or not.   This rather unexpected subtle discrepancy  contrasts  DFT with the traditional approaches  to the   various  non-Riemannian gravities. \\

\noindent The organization of the present  paper is as follows. \\
In the remaining of this Introduction,  to put the present work into context and  set up  notation,   we describe     DFT as  the  $\ODD$ completion of General Relativity along with  a relevant doubled string  action.    \\
In section~\ref{SECnbrn}, we review the  $(n,\brn)$ classification of the non-Riemannian DFT  geometries  from \cite{Morand:2017fnv}. \\  
In section~\ref{SECVP},  we revisit the variational principle in DFT and confirm   that the known  Euler--Lagrangian equations, or  `Einstein Double Field Equations'~(\ref{EDFE})  are  still  valid     for  non-Riemannian sectors. \\
In section~\ref{SECfix},   now keeping $(n,\brn)$ fixed, we reanalyze the variational principle and show that the full Einstein Double Equations are not necessarily implied  when $n\brn\neq0$.  We explain the discrepancy, and further propose  a non-Riemannian  differential tool kit as a `bookkeeping device' to expound the equations.\\
We conclude  in section~\ref{SECconclusion}, followed by Appendix~\ref{SECProof} \&      \ref{SECDerivation}. 

\subsection{Double Field Theory as the  $\ODD$ completion of General Relativity} 
While the initial motivation of Double Field Theory  was to reformulate supergravity in an $\ODD$ manifest manner~\cite{Siegel:1993xq,Siegel:1993th,Hull:2009mi,Hull:2009zb,Hohm:2010jy,Hohm:2010pp} (\cite{Aldazabal:2013sca,Berman:2013eva,Hohm:2013bwa} for  reviews), through subsequent  further developments~\cite{Jeon:2010rw,Jeon:2011cn,Hohm:2011si,Angus:2018mep},   DFT has evolved and  can be now   identified  as  a pure gravitational theory that  string theory seems to predict foremost\footnote{At least  formally let alone its phenomenological validity,~\textit{c.f.~}\cite{Ko:2016dxa,Park:2017snt}.}  and may differ from General Relativity as it is capable of describing non-Riemannian geometries~\cite{Morand:2017fnv}.  Specifically,   DFT  is the string theory based,  $\ODD$ completion of General Relativity (GR):  taking the   $\ODD$  symmetry   as the first principle,  DFT     geometrises  not merely the Riemannian metric but  the whole massless NS-NS sector of closed string as the fundamental  gravitational multiplet,  hence `completing' GR.  Further,  the $\ODD$ symmetry principle fixes its coupling to other superstring   sectors  (R-R~\cite{Rocen:2010bk,Hohm:2011zr,Hohm:2011dv,Jeon:2012kd}, R-NS~\cite{Jeon:2011vx},   and heterotic Yang-Mills~\cite{Hohm:2011ex,Hohm:2014sxa,Cho:2018alk}).   Having said that,       regardless of supersymmetry,   it can also    couple to    various     matter fields    which may appear  in lower dimensional effective field theories~\cite{Jeon:2011kp,Jeon:2011vx,Bekaert:2016isw}, just as GR does so. In particular, supersymmetric extensions have been completed    to the full (\textit{i.e.~}quartic)  order in fermions for ${D=10}$ cases powered by 
`1.5 formalism'~\cite{Jeon:2011sq,Jeon:2012hp}, and the  pure Standard Model without any extra physical degrees of freedom  can easily couple  to   DFT in an $\ODD$ symmetric manner~\cite{Choi:2015bga}. 

Schematically,  governed  by  the $\ODD$ symmetry  principle,  DFT  may couple to generic matter fields,   say collectively  $\Upsilon$, which should be  also in $\ODD$ representations:
\be
\int~\textstyle{\frac{1}{16\pi G}}\,e^{-2d}\So+\cL_{{{\rm{matter}}}}(\Upsilon, \na_{A}\Upsilon)\,.
\label{ACTION}
\ee
Here, $d$ is the $\ODD$ singlet DFT-dilaton,  $\So$ is the DFT scalar curvature, and   $\na_{A}\Upsilon$  denotes  the covariant derivative of a matter field.  
To manifest  the $\ODD$ symmetry, the action is equipped with an $\ODD$ invariant metric, 
\be
\cJ_{AB}={\tiny{\mathbf{{\left(\ba{cc}\bf{0}&\bf{1}\\\bf{1}&\bf{0}\ea\right)}\,}}},
\label{cJ}
\ee
which, with its inverse $\cJ^{AB}$, is going to be  always used to lower and raise the $\ODD$ vector indices (Latin capital letters). It splits     the   doubled coordinates   into two parts,  $x^{A}=(\tx_{\mu},x^{\nu})$ and    $\partial_{A}=(\tpartial^{\mu},\partial_{\nu})$. Note that the doubling of the coordinates is crucial to manifest the $\ODD$ symmetry in DFT.   Like  GR,  the  General Covariance  (DFT-diffeomorphisms)  of the action~(\ref{ACTION})   naturally  gives rise to the definitions of  the  $\ODD$ completions of the  Einstein curvature,  $G_{AB}$~\cite{Park:2015bza} and also  the  Energy-Momentum tensor,  $T_{AB}$~\cite{Angus:2018mep}, of which the former and the latter are respectively off-shell and on-shell conserved. Equating the two, they comprise   the  $\ODD$ completion of the Einstein field  equations, or the Einstein Double Field Equations (EDFEs)~\cite{Angus:2018mep,Park:2019hbc},
\be
G_{AB}=8\pi G\,T_{AB}\,.
\label{EDFE}
\ee
We summarize the basic geometrical notation of DFT   in Table~\ref{TableDFT}\footnote{The expression of $S_{AB}$ in Table~\ref{TableDFT} is newly   derived   from \cite{Jeon:2011cn} using 
$\Gamma_{ACD}\Gamma^{CBD}=\Gamma^{BCD}\Gamma_{CAD}
=\half\Gamma_{ACD}\Gamma^{BCD\,}$ and $
\Gamma_{CAD}\Gamma^{DBC}=\Gamma_{CAD}\Gamma^{CBD}
-\half\Gamma_{ACD}\Gamma^{BCD\,}$ which hold due to    the symmetric properties, $\Gamma_{[ABC]}=0$ and  $\Gamma_{A(BC)}=0$.
},  while the DFT-diffeomorphisms are generated by the so-called generalized Lie derivative~\cite{Siegel:1993th,Hohm:2010pp}: acting on a tensor  density with weight $\omega_{\scriptscriptstyle{T}}$,
\be
\delta_{\xi}T_{A_{1}\cdots A_{n}}=\hcL_{\xi}T_{A_{1}\cdots A_{n}}=
\xi^{B}\partial_{B}T_{A_{1}\cdots A_{n}}+\omegaT\partial_{B}\xi^{B}\,T_{A_{1}\cdots A_{n}}
+\sum_{j=1}^{n}(\partial_{A_{j}}\xi_{B}-\partial_{B}\xi_{A_{j}})T_{A_{1}\cdots A_{j-1}}{}^{B}{}_{A_{j+1}\cdots A_{n}}\,.
\label{gLie}
\ee
In particular, being a scalar density with weight one (${\omegaT=1}$),    the exponentiation $e^{-2d}$ is the integral measure of DFT.

\begin{center}
\begin{table}
\begin{tabular}{cc}
\hline
{Integral measure}
& $e^{-2d}$\quad(\textit{weight one scalar density})\\
 \begin{tabular}{c}
{Projectors}  \\
~
\end{tabular}
& $\ba{c}
P_{AB}=P_{BA}=\half (\cJ_{AB}+\cH_{AB})\,,\qquad\brP_{AB}=\brP_{BA}=\half (\cJ_{AB}-\cH_{AB})\\
P_{A}{}^{B}P_{B}{}^{C}=P_{A}{}^{C}\,,\qquad
\brP_{A}{}^{B}\brP_{B}{}^{C}=\brP_{A}{}^{C}\,,\qquad
P_{A}{}^{B}\brP_{B}{}^{C}=0
\ea$ \\ 
 \begin{tabular}{l}
{Christoffel symbols} \\
~
\end{tabular}
&\!\!\! ${\ba{lll}\Gamma_{CAB}&\!\!\!=&\!\!\!2\left(P\partial_{C}P\brP\right)_{[AB]}
+2\left({{\brP}_{[A}{}^{D}{\brP}_{B]}{}^{E}}-{P_{[A}{}^{D}P_{B]}{}^{E}}\right)\partial_{D}P_{EC}\\
{}&{}&\!\!\!\!\!\!\!\!\!\!\!\!\!\!\!\!\!\!\!\!\!\!
{{-4\left(\textstyle{\frac{1}{P_{M}{}^{M}-1}}P_{C[A}P_{B]}{}^{D}+\textstyle{\frac{1}{\brP_{M}{}^{M}-1}}\brP_{C[A}\brP_{B]}{}^{D}\right)\!}}\left(\partial_{D}d+(P\partial^{E}P\brP)_{[ED]}\right)\ea}$ \\  
{Covariant derivatives}& $~
P_{A}{}^{C}\brP_{B}{}^{D}\na_{C}V_{D}$, \qquad $\brP_{A}{}^{C}P_{B}{}^{D}\na_{C}V_{D}$,\qquad$P^{AB}\na_{A}V_{B}$,\qquad$\brP^{AB}\na_{A}V_{B}$\\
{Semi-covariant derivative}&$~\na_{C}V_{D}=\partial_{C}V_{D}
-\omega_{\scriptscriptstyle{V}}\Gamma^{E}{}_{EC}V_{D}
+\Gamma_{CD}{}^{E}V_{E}$\\
{Compatibility}&$\na_{C}P_{AB}=\na_{C}\brP_{AB}=\na_{C}\cJ_{AB}=0$,\qquad $\na_{C}d=-\half e^{2d}\na_{C}\left(e^{-2d}\right)=0$\\
{Scalar curvature}& $\So=\cH^{AB}S_{AB}$\\
{Ricci curvature}&$(PS\brP)_{AB}=P_{A}{}^{C}\brP_{B}{}^{D}S_{CD}$\\
{Einstein curvature}&  $G_{AB}=4P_{[A}{}^{C}\brP_{B]}{}^{D}S_{CD}-\half\cJ_{AB}\So$\\
{Semi-covariant curvature}&$~S_{AB}=2\partial_{A}\partial_{B}d-e^{2d\,}\partial_{C}\left(e^{-2d\,}\Gamma_{(AB)}{}^{C}\right)+
\half\Gamma_{ACD}\Gamma_{B}{}^{CD}-\half\Gamma_{CDA}\Gamma^{CD}{}_{B}$\\
{Variational\, property}&$\delta S_{AB}=\na_{[A}\delta\Gamma_{C]B}{}^{C}+\na_{[B}\delta\Gamma_{C]A}{}^{C}$\\
{Energy-Momentum tensor}& 
$T^{AB}=e^{2d}\left(8
\brP^{[A}{}_{C}P^{B]}{}_{D\,}\frac{\delta\cL_{{\rm{matter}}}}{\delta\cH_{CD}}-\frac{1}{2}\cJ^{AB\,}\frac{\delta\cL_{{\rm{matter}}}}{\delta d}\right)$\\
{Conservation}&$\na_{A}G^{AB}=0$\quad (\textit{off-shell})\,,\qquad 
$\na_{A}T^{AB}=0$\quad (\textit{on-shell})\\
{EDFEs}& $G_{AB}=8\pi G\,T_{AB}$\\
\hline
\end{tabular}
\caption{{\bf{Geometric notation for DFT.}} For latest  exposition see \textit{e.g.~}section 2 of  \cite{Angus:2018mep}.  }
\label{TableDFT}
\end{table}
\end{center}

It is noteworthy and relevant to this work that, all the geometrical notation of  the  covariant derivative, $\na_{A}$, and    the  curvatures, $\So,G_{AB}$,    can be  constructed  strictly  in terms of  $\ODD$ covariant field variables, notably     the $\ODD$ invariant   DFT-dilaton, $d$,  and  the $\ODD$ covariant   DFT-metric, $\cH_{AB}$ (``generalized metric"),  or more powerfully $\ODD$ covariant  DFT-vielbeins, without necessarily referring to  conventional, undoubled $\ODD$ breaking supergravity variables.  Similarly, a doubled string action can be constructed in terms of  $\ODD$ covariant objects as we review below.

\subsection{Doubled but at the same time gauged string action} 
One of the characteristics  of DFT is  the   imposition of  the  `section condition': acting on arbitrary functions  in DFT, say $\Phi_{r}$, and their  products like $\Phi_{s}\Phi_{t}$,  the $\ODD$ invariant Laplacian should vanish
\be
\ba{llll}
\partial_{A}\partial^{A}=0&~~~:&~~~
\partial_{A}\partial^{A}\Phi_{r}=0\,,~~~&~~~
\partial_{A}\Phi_{s}\partial^{A}\Phi_{t}=0\,.
\ea
\ee
We remind  the reader that the $\ODD$   indices are raised with $\cJ^{AB}$. 
Upon  imposing  the section condition, the generalized Lie derivative~(\ref{gLie}) is  closed by commutators~\cite{Siegel:1993th,Hohm:2010pp},
\be
\ba{ll}
\left[\hcL_{\zeta},\hcL_{\xi}\right]=\hcL_{\left[\zeta,\xi\right]_{\rm{C}}}\,,
\quad&\qquad
\left[\zeta,\xi\right]^{M}_{\rm{C}}= \zeta^{N}\partial_{N}\xi^{M}-\xi^{N}\partial_{N}\zeta^{M}+\half \xi^{N}\partial^{M}\zeta_{N}-\half \zeta^{N}\partial^{M}\xi_{N}\,.
\ea
\label{closeda}
\ee

The section condition  is mathematically equivalent to the following    translational invariance~\cite{Park:2013mpa,Lee:2013hma},
\be
\ba{ll}
\Phi_{r}(x)=\Phi_{r}(x+\Delta)\,,\qquad&\qquad\Delta^{A}\partial_{A}=0\,,
\ea
\label{invariance}
\ee
where the shift parameter, $\Delta^{A}$, is \textit{derivative-index-valued}, meaning that its superscript index should be identifiable as a derivative index, for example  $\Delta^{A}=\Phi_{s}\partial^{A}\Phi_{t\,}$.  This insight on the section condition  may suggest that the doubled coordinates of DFT are in fact gauged by an equivalence relation,
\be
\ba{ll}
x^{A}\quad\sim\quad x^{A}+\Delta^{A}\,,\qquad&\qquad \Delta^{A}\partial_{A}=0\,.
\ea
\label{gauging}
\ee
Each gauge orbit, \textit{i.e.~}equivalence class, represents a single physical point.  As a matter of fact  in DFT,  the usual infinitesimal one-form of coordinates, $\rmd x^{A}$, is not DFT-diffeomorphism covariant, 
\be
\delta(\rmd x^{A})=\rmd(\delta x^{A})=\rmd\xi^{A}=\rmd x^{B}\partial_{B}\xi^{A}\neq\rmd x^{B}(\partial_{B}\xi^{A}-\partial^{A}\xi_{B})\,.
\ee
However,  if we  gauge the one-form  by   introducing     a derivative-index-valued connection,   we can have  a  DFT-diffeomorphism covariant  one-form,  provided that the gauge potential transforms appropriately,
\be
\ba{llll}
\rmD x^{A}=\rmd x^{A}-\cA^{A}\,,\quad&\quad \cA^{A}\partial_{A}=0\,,\quad&\quad
\delta(\rmD x^{A})=\rmD x^{B}(\partial_{B}\xi^{A}-\partial^{A}\xi_{B})\,,\quad&\quad
\delta \cA^{A}=\rmD x^{B}\partial^{A}\xi_{B}\,.
\ea
\label{rmDxcA}
\ee
It is also a singlet of the coordinate gauge symmetry~(\ref{gauging}):    $\delta x^{A}=\Delta^{A},\, \delta\cA^{A}=\rmd\Delta^{A},\, \delta(\rmD x^{A})=0$.  The gauged one-form then naturally  allows   to construct a  perfectly symmetric doubled   string action~\cite{Hull:2006va},\cite{Lee:2013hma},
\be
{\textstyle{\frac{1}{4\pi\alpha^{\prime}}}}{\displaystyle{\int}}\rmd^{2}\sigma~\left[
-\half\sqrt{-h}h^{\alpha\beta}\rmD_{\alpha}x^{A}\rmD_{\beta}x^{B}\cH_{AB}
-\epsilon^{\alpha\beta}\rmD_{\alpha}x^{A}\cA_{\beta A}\right]\,,
\label{stringaction}
\ee
which enjoys symmetries like   global  $\ODD$,   target spacetime DFT-diffeomorphisms,  worldsheet  diffeomorphisms,  Weyl symmetry, and the coordinate gauge symmetry.\footnote{See  also \cite{Park:2016sbw} for    Green--Schwarz  doubled superstring, \cite{Ko:2016dxa} for  doubled point particle, and \cite{Arvanitakis:2017hwb,
Arvanitakis:2018hfn} for `exceptional' extensions.}  All the background information is  encoded in the DFT-metric, $\cH_{AB}\,$.

\section{Review of \cite{Morand:2017fnv}: Classification of the non-Riemannian  DFT  geometries  \label{SECnbrn}} 
The  section condition  can be generically solved, up to $\ODD$ rotations,      by enforcing  the   tilde coordinate independency: $\tpartial^{\mu}\equiv0~\Rightarrow~ \partial_{A}\partial^{A}=2\partial_{\mu}\tpartial^{\mu}\equiv0$.   Choosing  $\Delta^{A}=c_{\mu}\partial^{A}x^{\mu}=(c_{\mu},0)$ for (\ref{gauging}) and similarly  $\cA^{A}=A_{\mu}\partial^{A}x^{\mu}=(A_{\mu},0)$,  we note that   
 the tilde coordinates are indeed gauged:  $(\tx_{\mu},x^{\nu})\,\sim\,(\tx_{\mu}+c_{\mu},x^{\nu})$, $\rmD x^{A}=(\rmd\tx_{\mu}-A_{\mu},\rmd x^{\nu})$. With respect to this choice of the section, the well-known  parametrization of  the   DFT-metric and the  DFT-dilaton in terms of the conventional  massless NS-NS   field  variables~\cite{Giveon:1988tt,Duff:1989tf},
\be
\ba{ll}
\cH_{AB}=\left(\ba{cc}g^{\mu\nu}&-g^{\mu\sigma}B_{\sigma\lambda}\\
B_{\kappa\rho}g^{\rho\nu}& g_{\kappa\lambda}-B_{\kappa\rho}g^{\rho\sigma}B_{\sigma\lambda}\ea\right),
\qquad&\qquad e^{-2d}=e^{-2\phi}\sqrt{\left|g\right|}\,,
\ea
\label{Riemannian}
\ee
reduces DFT to  supergravity. In this case,  the single expression of the EDFEs~(\ref{EDFE}) unifies  all the   equations of motion of the  three  fields, $\{g_{\mu},B_{\mu\nu},\phi\}$.  Further,  after Gaussian integration of the auxiliary gauge potential,  $A_{\mu}$,  the doubled-yet-gauged  string action~(\ref{stringaction})   reproduces the standard undoubled string action.  

Yet, this is not the full story. The above parametrization~(\ref{Riemannian}) is merely  one  particular   solution to the defining relations of the DFT-metric:
\be
\ba{ll}
\cH_{AB}=\cH_{BA}\,,\qquad&\qquad\cH_{A}{}^{C}\cH_{B}{}^{D}\cJ_{CD}=\cJ_{AB}\,.
\ea
\label{defining}
\ee
DFT  and the doubled-yet-gauged string action   work well,  provided   these conditions are fulfilled.   For example,   instead of (\ref{Riemannian}), we may let the DFT-metric coincide with the $\ODD$ invariant metric, 
\be
\cH_{AB}={\tiny{\mathbf{{\left(\ba{cc}\bf{0}&\bf{1}\\\bf{1}&\bf{0}\ea\right)}\,}}},
\label{MAX}
\ee
such that $\cH_{A}{}^{B}=\delta_{A}{}^{B}$. This is a   vacuum solution to  DFT, or to the `matter-free'  EDFEs, $G_{AB}=0$~(\ref{EDFE}), which is very special  in  several aspects.    Firstly, compared with (\ref{Riemannian}),   there cannot be any associated Riemannian metric~$g_{\mu\nu}$ and hence it does not allow  any conventional or Riemannian interpretation at all. It is maximally non-Riemannian.  Secondly, it is  fully $\ODD$ symmetric, being one of  the two most symmetric vacua  of  DFT,  $\cH_{AB}=\pm\cJ_{AB}$.  Thirdly,  it  is  moduli-free since  it does not admit  any   infinitesimal fluctuation, $\delta\cH_{AB}=0$~\cite{Cho:2018alk}.\footnote{Put $\cH_{A}{}^{B}=\delta_{A}{}^{B}$ in (\ref{twoCON}).} And lastly but not leastly,   upon this background,   the  auxiliary gauge potential, $A_{\mu}$, appears  linearly rather than quadratically in the doubled-yet-gauged string action~(\ref{stringaction}).   Consequently it serves as  a Lagrange multiplier to prescribe that all the untilde target spacetime coordinates should be chiral~\cite{Lee:2013hma} (\textit{c.f.~}\cite{Siegel:2015axg,Casali:2017mss}),
\be
\partial_{\alpha}x^{\mu}+\textstyle{\frac{1}{\sqrt{-h}}}\epsilon_{\alpha}{}^{\beta}\partial_{\beta}x^{\mu}=0\,.
\ee   
An intriguing insight from \cite{Berman:2019izh} is then  that, the  usual supergravity  fields in (\ref{Riemannian}) would be  the Nambu--Goldstone modes of the  perfectly  $\ODD$ symmetric  vacuum~(\ref{MAX}).

Given the Riemannian and  maximally non-Riemannian backgrounds, (\ref{Riemannian}) \textit{v.s.}~(\ref{MAX}), one may wonder about the existence of  more generic non-Riemannian geometries (\textit{c.f.~}\cite{Lee:2013hma,Ko:2015rha} for other examples  and also \cite{Malek:2013sp} for  `timelike' duality rotations). This question was answered in \cite{Morand:2017fnv}: the most general solutions to the defining properties  of the DFT-metric~(\ref{defining}) can be  classified by two non-negative integers, $(n,\brn)$,
\be
\cH_{AB}=\left(\ba{cc}H^{\mu\nu}&
-H^{\mu\sigma}B_{\sigma\lambda}+Y_{i}^{\mu}X^{i}_{\lambda}-
\brY_{\bri}^{\mu}\brX^{\bri}_{\lambda}\\
B_{\kappa\rho}H^{\rho\nu}+X^{i}_{\kappa}Y_{i}^{\nu}
-\brX^{\bri}_{\kappa}\brY_{\bri}^{\nu}\quad&~~
~~K_{\kappa\lambda}-B_{\kappa\rho}H^{\rho\sigma}B_{\sigma\lambda}
+2X^{i}_{(\kappa}B_{\lambda)\rho}Y_{i}^{\rho}
-2\brX^{\bri}_{(\kappa}B_{\lambda)\rho}\brY_{\bri}^{\rho}
\ea\right),
\label{cHFINAL}
\ee
where $i,j=1,2,\cdots, n$,  $\bri,\brj=1,2,\cdots,  \brn$ \,and\,   $0\leq n+\brn\leq D$.
\begin{itemize}
\item[\textit{(i)}] While the $B$-field is skew-symmetric as usual, $H^{\mu\nu}$ and $K_{\mu\nu}$ are symmetric tensors  
whose kernels are spanned by linearly independent vectors, $\big\{X^{i}_{\mu},\brX^{\bri}_{\nu}\big\}$  and $\big\{Y_{j}^{\mu},\brY^{\nu}_{\brj}\big\}$, respectively,
\be
\ba{llll}
H^{\mu\nu}X^{i}_{\nu}=0\,,\quad&\quad
H^{\mu\nu}\brX^{\bri}_{\nu}=0\,,~\quad&~\quad
K_{\mu\nu}Y_{j}^{\nu}=0\,,\quad&\quad
K_{\mu\nu}\brY_{\brj}^{\nu}=0\,.
\ea
\label{HXX}
\ee
\item[\textit{(ii)}]
A completeness relation must be satisfied 
\be
H^{\mu\rho}K_{\rho\nu}
+Y_{i}^{\mu}X^{i}_{\nu}+\brY_{\bri}^{\mu}\brX^{\bri}_{\nu}
=\delta^{\mu}{}_{\nu}\,.
\label{COMP}
\ee
\end{itemize}
From the linear independency of the zero-eigenvectors,  $\left\{X^{i}_{\mu}, \brX^{\bri}_{\nu}\right\}$,   orthogonal/algebraic  relations  follow
\be
\ba{lllll}
Y^{\mu}_{i}X_{\mu}^{j}=\delta_{i}{}^{j}\,,\quad&\,
\brY^{\mu}_{\bri}\brX_{\mu}^{\brj}=\delta_{\bri}{}^{\brj}\,,\quad&\,
Y^{\mu}_{i}\brX_{\mu}^{\brj}=
\brY^{\mu}_{\bri}X_{\mu}^{j}=0\,,\quad&\,
H^{\rho\mu}K_{\mu\nu}H^{\nu\sigma}=H^{\rho\sigma}\,,\quad&\,
K_{\rho\mu}H^{\mu\nu}K_{\nu\sigma}=K_{\rho\sigma}\,.
\ea
\label{complete}
\ee
Intriguingly,  the $B$-field  (hence `Courant algebra') is universally present regardless of the values of $(n,\brn)$, and contributes to the DFT-metric through  an $\ODD$  adjoint action:
\be
\cH_{AB}=\cB_{A}{}^{C}\cB_{B}{}^{D}\mcH_{CD}\,,
\label{DECOMP}
\ee
where $\mcH$  corresponds to  the  `$B$-field-free' DFT-metric,
\be
\mcH_{AB}=\left(\ba{cc}H~&~
Y_{i}(X^{i})^{T}-\brY_{\bri}(\brX^{\bri})^{T}\\
X^{i}(Y_{i})^{T}-\brX^{\bri}(\brY_{\bri})^{T}\quad~&~\quad K
\ea\right),
\label{mcH}
\ee
and $\cB$ is  an $\ODD$ element containing the $B$-field,
\be
\ba{ll}
\cB_{A}{}^{B}=\left(\ba{cc}\delta^{\mu}{}_{\sigma}&0\\B_{\rho\sigma}&~\delta_{\rho}{}^{\tau}\ea\right),
\qquad&\qquad
\cB_{A}{}^{C}\cB_{B}{}^{D}\cJ_{CD}=\cJ_{AB}\,.
\ea
\label{cB}
\ee
It is also worth while to note the `vielbeins' or `square-roots' of $K_{\mu\nu}$ and $H^{\mu\nu}$\,: 
\be
\ba{llll}
K_{\mu\nu}=K_{\mu}{}^{a}K_{\nu}{}^{b}\eta_{ab}\,,\qquad&\qquad
H^{\mu\nu}=H^{\mu}{}_{a}H^{\nu}{}_{b}\eta^{ab}\,,\qquad&\qquad
K_{\mu}{}^{a}H^{\mu}{}_{b}=\delta^{a}{}_{b}\,,\qquad&\qquad
K_{\mu a}H^{\nu a}=K_{\mu\rho}H^{\rho\nu}\,,
\ea
\label{KHroots}
\ee
where  $a,b$ are $(D-n-\brn)$-dimensional  indices subject to a flat metric, say $\eta_{ab}\,$,  whose signature is  arbitrary. Essentially,  $\big\{K_{\mu}{}^{a},X_{\mu}^{i},\brX_{\mu}^{\bri}\big\}$ form a $D\times D$ invertible square matrix whose inverse is given by   $\big\{H^{\mu}{}_{a},Y^{\mu}_{i},\brY^{\mu}_{\bri}\big\}$ as
\be
K_{\mu}{}^{a}H^{\nu}{}_{a}+X_{\mu}^{i}Y^{\nu}_{i}+\brX_{\mu}^{\bri}\brY^{\nu}_{\bri}=\delta_{\mu}{}^{\nu}\,.
\label{vcomp}
\ee
In fact, the analysis of the DFT-vielbeins corresponding to  the $(n,\brn)$ DFT-metric~(\ref{cHFINAL}) carried out in \cite{Morand:2017fnv} shows  that  the local Lorentz symmetry group, \textit{i.e.~}spin group is
\be
\Spin(t+n,s+n)\times\Spin(s+\brn,t+\brn)\,.
\label{LLS}
\ee
Here  $(t,s)$ is the arbitrary signature of $\eta_{ab}$ or the nontrivial signature of $H^{\mu\nu}$ and  $K_{\mu\nu}$ satisfying   $t+s+n+\brn=D$.  Of course, once the spin group of any  given theory is specified, it is  fixed   once and  for all. Thus, each sum,  $t+n$, $s+n$, $s+\brn$, and $t+\brn$, should be  constant.  For example,  the Minkowskian $D=10$ maximally  supersymmetric DFT~\cite{Park:2016sbw} and the doubled-yet-gauged Green-Schwarz superstring action~\cite{Jeon:2012hp}, both having  the  local Lorentz group  of $\Spin(1,9)\times\Spin(9,1)$,  can  accommodate $(0,0)$ Riemannian and $(1,1)$ non-Riemannian sectors only (see \cite{Sakatani:2019jgu} for examples of  supersymmetric non-Riemannian backgrounds).  Nevertheless,  we may readily  relax the Majorana--Weyl condition therein~\cite{Jeon:2012hp,Park:2016sbw}  and impose  the Weyl condition  only on spinors, such that  the  local Lorentz  group can take  any of  $\Spin(\hat{t},\hat{s})\times\Spin(\hat{s},\hat{t})$ with $\hat{t}+\hat{s}=10$. The allowed non-Riemannian geometries will be  then $(n,n)$ types with ${n=\brn}$ running from zero to ${\textrm{min}}(\hat{t},\hat{s})$~\cite{Morand:2017fnv}. On the other hand, bosonic DFT does not care about  spin groups   and  hence  
should be free from such  constraints. It can admit more generic    $(n,\brn)$ non-Riemannian geometries.

Crucially, the $(n,\brn)$ parametrization of the DFT-metric~(\ref{cHFINAL})  possesses  two local symmetries, namely $\GL(n)\times\GL(\brn)$ rotations and Milne-shift transformations.   The  $\GL(n)\times\GL(\brn)$ symmetry rotates    the $i,j,\cdots$ and $\bri,\brj,\cdots$ indices of the component fields:  with  infinitesimal  local  parameters, $w_{i}{}^{j}$ and  $\brw_{\bri}{}^{\brj}$,
\be
\ba{cccc}
\deltaw X^{i}_{\mu}=X^{j}_{\mu\,}w_{j}{}^{i}\,,\quad&\quad
\deltaw Y_{i}^{\mu}=-w_{i}{}^{j\,}Y_{j}^{\mu}\,,\quad&\quad
\deltaw \brX^{\bri}_{\mu}=\brX^{\brj}_{\mu\,}\brw_{\brj}{}^{\bri}\,,\quad&\quad
\deltaw \brY_{\bri}^{\mu}=-\brw_{\bri}{}^{\brj\,}\brY_{\brj}^{\mu}\,,\\
\deltaw d=0\,,\quad&\quad\deltaw H^{\mu\nu}=0\,,\quad&\quad
\deltaw K_{\mu\nu}=0\,,\quad&\quad\deltaw B_{\mu\nu}=0\,.
\ea
\label{deltaw}
\ee
The  Milne-shift symmetry generalizes  the so-called  `Galilean  boost' in the Newtonian  gravity literature~\cite{Milne:1934,Duval:1993pe}. It acts  with  infinitesimal  local  parameters, $V_{\mu i}$ and  $\brV_{\mu\bri}$,\footnote{Through exponentiations,    finite Milne-shift  transformations can be achieved, which turn out to   get  truncated at finite orders, for example $e^{\deltaM}Y_{i}^{\mu}=Y_{i}^{\mu}+H^{\mu\nu}V_{\nu i}\,$. See Eq.(2.16) of \cite{Morand:2017fnv} for the full list.}
\be
\ba{c}
\deltaM Y^{\mu}_{i}=H^{\mu\nu}V_{\nu i}\,,\quad\qquad\qquad
\deltaM\brY_{\bri}^{\mu}=H^{\mu\nu}\brV_{\nu\bri}\,,\\
\deltaM K_{\mu\nu}=-2X^{i}_{(\mu}K_{\nu)\rho}H^{\rho\sigma}V_{\sigma i}-2\brX^{\bri}_{(\mu}K_{\nu)\rho}H^{\rho\sigma}\brV_{\sigma\bri}\,,\\
\deltaM
B_{\mu\nu}=
-2X^{i}_{[\mu}V_{\nu]i}+2\brX^{\bri}_{[\mu}\brV_{\nu]\bri}
+2X^{i}_{[\mu}\brX^{\bri}_{\nu]}\left(Y_{i}^{\rho}\brV_{\rho\bri}
+\brY_{\bri}^{\rho}V_{\rho i}\right),\\
\deltaM d=0\,,\quad\qquad
\deltaM X^{i}_{\mu}=0\,,\quad\qquad
\deltaM\brX^{\bri}_{\mu}=0\,,\quad\qquad \deltaM H^{\mu\nu}=0\,.
\ea
\label{MS}
\ee
Remarkably, both  transformations, (\ref{deltaw}) and  (\ref{MS}),  leave  the DFT-metric  invariant, 
\be
\ba{ll}
\deltaw \cH_{AB}=0\,,\qquad&\qquad\deltaM\cH_{AB}=0\,,
\ea
\ee
as  the two local symmetries are  actually   parts of the underlying  local Lorentz symmetries~(\ref{LLS}).

Upon the $(n,\brn)$ background, the doubled-yet-gauged worldsheet string action~(\ref{stringaction}) reduces to
\be
{\textstyle{\frac{1}{2\pi\alpha^{\prime}}}}{\displaystyle{\int}}\rmd^{2}\sigma~\left[
-\half\sqrt{-h}h^{\alpha\beta}\partial_{\alpha}x^{\mu}\partial_{\beta}x^{\nu}
K_{\mu\nu}
+\half\epsilon^{\alpha\beta}\partial_{\alpha}x^{\mu}
\partial_{\beta}x^{\nu}B_{\mu\nu}+\half\epsilon^{\alpha\beta}
\partial_{\alpha}\tx_{\mu}\partial_{\beta}x^{\mu}\right]\,,
\ee
which should be  supplemented by the chiral and anti-chiral constraints  over the $n$ and $\brn$ directions,
\be
\ba{ll}
X^{i}_{\mu}\left(\partial_{\alpha}x^{\mu}+\textstyle{\frac{1}{\sqrt{-h}}}\epsilon_{\alpha}{}^{\beta}\partial_{\beta}x^{\mu}\right)=0\,,
\quad&\quad
\brX^{\bri}_{\mu}\left(\partial_{\alpha}x^{\mu}-\textstyle{\frac{1}{\sqrt{-h}}}\epsilon_{\alpha}{}^{\beta}\partial_{\beta}x^{\mu}\right)=0\,.
\ea
\ee
These constraints are prescribed by the  integrated-out  auxiliary gauge potential~$\cA^{A}$~(\ref{rmDxcA}).\\
~\\
\noindent\textbf{Comment 1.}  Matching  with the content of the  non-Riemannian component fields, 
\be
\{H^{\mu\nu},K_{\rho\sigma},X_{\mu}^{i},\brX_{\nu}^{\bri},Y^{\rho}_{j},\brY^{\sigma}_{\brj},B_{\mu\nu}\}\,,
\ee 
and   the undoubled  string worldsheet action resulting  from  (\ref{stringaction}),  one can identify   the original  Newton--Cartan~\cite{Cartan:1923zea,Kuenzle:1972zw,Duval:1984cj} as $(1,0)$,  Stringy Newton--Cartan~\cite{Andringa:2012uz} as $(1,1)$,  Carroll~\cite{Henneaux79,Duval:2014uoa}   as $(D{-1},0)$, and   Gomis--Ooguri~\cite{Gomis:2000bd}  as $(1,1)$: see \cite{Morand:2017fnv,Berman:2019izh,Blair:2019qwi} for the details of the identifications.     Further,   the  isometry of the  $(1,1)$ flat DFT-metric   matches with the non-relativistic symmetry algebra such as Bargmann algebra~\cite{Ko:2015rha}, while the notion of T-duality persists to make sense  in the  non-relativistic string theory~\cite{Bergshoeff:2018yvt}.   These all seem to suggest that DFT may be the \textit{home}, \textit{i.e.~}the  unifying framework,  to describe   various known as well as yet-unknown non-Riemannian gravities.\footnote{Similarly,  inequivalent parametrizations of the DFT-vielbeins, or U-duality-covariant generalized metric,  correspond to  the conventional distinctions between  IIA and IIB~\cite{Jeon:2012hp,Park:2016sbw}, or IIB and M-``theories"~\cite{Blair:2013gqa}.}
 Having said that there are also a few novel ingredients  from DFT,  such as the local $\GL(n)\times\GL(\brn)$ symmetry~(\ref{deltaw}), the notion of  `Milne-shift covariance' as we shall discuss  below~(\ref{Mcov}), (\ref{Mcov2}),  and the very existence of the DFT-dilaton of which the exponentiation, $e^{-2d}$, gives the integral measure in DFT being a scalar density with weight one, 
\be
\delta_{\xi}d=-\half e^{2d}\cL_{\xi}\left(e^{-2d}\right)=-\half e^{2d}\partial_{\mu}\left(\xi^{\mu}e^{-2d}\right)
=\xi^{\mu}\partial_{\mu}d-\half\partial_{\mu}\xi^{\mu}\,.
\ee

\noindent\textbf{Comment 2.} It is worth while to generalize the decomposition~(\ref{DECOMP}) to an arbitrary DFT tensor,
\be
\ba{ll}
\mT_{A_{1}A_{2}\cdots A_{n}}:=(\mcB^{-1})_{A_{1}}{}^{B_{1}}(\mcB^{-1})_{A_{2}}{}^{B_{2}}\cdots(\mcB^{-1})_{A_{n}}{}^{B_{n}}T_{B_{1}B_{2}\cdots B_{n}}\,,\quad&~~
T_{A_{1}\cdots A_{n}}=\mcB_{A_{1}}{}^{B_{1}}\cdots\mcB_{A_{n}}{}^{B_{n}}\mT_{B_{1}\cdots B_{n}}\,.
\ea
\label{gDCMP}
\ee
Under diffeomorphisms,  while the DFT tensor $T_{A_{1}\cdots A_{n}}$ is surely  subject to the generalized Lie derivative~(\ref{gLie}),   the circled  quantity,   $\mT_{A_{1}\cdots A_{n}}$,  is now  governed by the  undoubled ordinary   Lie derivative which can be conveniently   obtained as the truncation of  the  generalized Lie derivative  by  choosing   the section, $\tpartial^{\mu}\equiv0$,  and setting the parameter, $\xi^{A}=(0,\xi^{\mu})$ as  $\tilde{\xi}_{\nu}\equiv0$: 
\be
\delta_{\xi}\mT_{A_{1}\cdots A_{n}}=\cL_{\xi}\mT_{A_{1}\cdots A_{n}}=
\xi^{\mu}\partial_{\mu}\mT_{A_{1}\cdots A_{n}}+\omegaT\partial_{\mu}\xi^{\mu}\,\mT_{A_{1}\cdots A_{n}}
+\sum_{j=1}^{n}(\partial_{A_{j}}\xi_{B}-\partial_{B}\xi_{A_{j}})\mT_{A_{1}\cdots A_{j-1}}{}^{B}{}_{A_{j+1}\cdots A_{n}}\,.
\label{oLie}
\ee
Further,   by construction, a DFT tensor  is  Milne-shift  \textit{invariant}. Yet,  the circled one is Milne-shift  \textit{covariant} in the following manner,
\be
\ba{ll}
\deltaM T_{A_{1}\cdots A_{n}}=0\,,\qquad&\qquad
\deltaM \mT_{A_{1}\cdots A_{n}}=\dis{\sum_{j=1}^{n}}-\deltaM\cB_{A_{j}}{}^{B}\mT_{A_{1}\cdots A_{j-1}BA_{j+1}\cdots A_{n}}\,.
\ea
\label{Mcov}
\ee
Explicitly,  for a DFT vector, $V_{A}=\cB_{A}{}^{B}\mV_{B}$, we have (\textit{c.f.~}\cite{Jeon:2011kp,Hull:2014mxa}) 
\be
\ba{l}
\delta_{\xi}\mV_{A}=
\left(\ba{c}\delta_{\xi}\mV^{\mu}\\\delta_{\xi}\mV_{\nu}\ea\right)=
\left(\ba{c}\cL_{\xi}\mV^{\mu}\\\cL_{\xi}\mV_{\nu}\ea\right)
=\xi^{\rho}\partial_{\rho}\mV_{A}+\omega_{\scriptscriptstyle{V}}\partial_{\rho}\xi^{\rho}\,\mV_{A}
+(\partial_{A}\xi_{B}-\partial_{B}\xi_{A})\mV^{B}
=\cL_{\xi}\mV_{A}\,,\\
\deltaM\mV_{A}=
\left(\ba{c}\deltaM\mV^{\mu}\\\deltaM\mV_{\nu}\ea\right)=
\left(\ba{c}0\\-\deltaM B_{\nu\rho}\mV^{\rho}\ea\right)
=-\deltaM\cB_{A}{}^{B}\mV_{B}\,.
\ea
\label{xMV}
\ee
That is to say,   the circled quantities, $\mT_{A_{1}\cdots A_{n}}$, $\mV_{A}$, are    `$B$-field free', subject to the ordinary Lie derivative, and  Milne-shift covariant rather than invariant. More specifically, the undoubled  lower  Greek indices are Milne-shift covariant, while the upper ones are invariant: from (\ref{mcH}),   (\ref{MS}),  (\ref{xMV}), 
\be
\ba{l}
\deltaM\mV_{\nu}=-\deltaM B_{\nu\rho}\mV^{\rho}\,,\\
\deltaM\mV^{\mu}=0\,,\\
\deltaM K_{\mu\nu}=\deltaM\mcH_{\mu\nu}
=-\deltaM B_{\mu\rho}\mcH^{\rho}{}_{\nu}-\deltaM B_{\nu\rho}\mcH_{\mu}{}^{\rho}
=-\deltaM B_{\mu\rho}(Y_{i}^{\rho}X^{i}_{\nu}-\brY_{\bri}^{\rho}\brX^{\bri}_{\nu})-\deltaM B_{\nu\rho}(Y_{i}^{\rho}X^{i}_{\mu}-\brY_{\bri}^{\rho}\brX^{\bri}_{\mu})\,,\\
\deltaM (Y_{i}^{\mu}X^{i}_{\nu}-\brY_{\bri}^{\mu}\brX^{\bri}_{\nu})=
\deltaM\mcH^{\mu}{}_{\nu}
=-\deltaM B_{\nu\rho}\mcH^{\mu\rho}=\deltaM\mcH_{\nu}{}^{\mu}=-\deltaM B_{\nu\rho}\mcH^{\rho\mu}
=-\deltaM  B_{\nu\rho}H^{\mu\rho}\,,\\
\deltaM H^{\mu\nu}=\deltaM \mcH^{\mu\nu}=0\,.
\ea
\label{Mcov2}
\ee
For consistency, we also note for the $\ODD$ invariant metric, 
\be
\ba{ll}
\cJ_{AB}=\mcJ_{AB}\,,\qquad&\qquad
\deltaM\mcJ_{AB}=-\deltaM \cB_{A}{}^{C}\mcJ_{CB}-\deltaM \cB_{B}{}^{C}\mcJ_{AC}=0\,.
\ea
\ee

\newpage

\section{Variational  Principle around  non-Riemannian backgrounds \label{SECVP}}
Here  we revisit with care   the variational principle  for  a general  DFT action coupled to  matter~(\ref{ACTION}) especially around  non-Riemannian backgrounds. While the variations of the matter fields lead to their own Euler--Lagrange  equations of motion,  the variations of the DFT-metric and the DFT-dilaton give~\cite{Angus:2018mep}
\be
\delta\dis{\int}\textstyle{\frac{1}{16\pi G}}\,e^{-2d}\So+\cL_{{{\rm{matter}}}}
=\textstyle{\frac{1}{16\pi G}}\!\dis{\int}e^{-2d}\Big[\delta\cH_{AB}\big\{
(PG\brP)^{AB}-8\pi G(PT\brP)^{AB}\big\}+\textstyle{\frac{2}{D}}\delta d(G_{A}{}^{A}-8\pi GT_{A}{}^{A})\Big]\,.
\label{AV}
\ee
Here $G_{AB}$ and $T_{AB}$ are respectively   the stringy  or $\ODD$ completions of the Einstein curvature~\cite{Park:2015bza} and the  Energy-Momentum tensor~\cite{Angus:2018mep}, as summarized in Table~\ref{TableDFT}. The above result  is easy to obtain  once  we  neglect  a boundary contribution arising  from a total derivative~\cite{Jeon:2011cn}:
\be
\textstyle{\frac{1}{16\pi G}}e^{-2d\,}\cH^{AB}\delta S_{AB}=\partial_{A}\left(e^{-2d}\textstyle{\frac{1}{8\pi G}}\cH^{B[A}\delta\Gamma_{CB}{}^{C]}\right)\,,
\label{POWER}
\ee
and take into account a well-known identity which  the infinitesimal variation of the DFT-metric should satisfy~\cite{Hohm:2010pp,Jeon:2010rw,Berkeley:2014nza}, 
\be
\delta\cH_{AB}=2P_{(A}{}^{C}\brP_{B)}{}^{D}\delta\cH_{CD}\,.
\label{deltacH0}
\ee
Eq.(\ref{POWER}) holds due to the nice variational  property of the semi-covariant curvature,  $\delta S_{AB}=\na_{[A}\delta\Gamma_{C]B}{}^{C}+\na_{[B}\delta\Gamma_{C]A}{}^{C}$, and the compatibility of the derivative, ${\na_{A}\cJ_{BC}=0}$, ${\na_{A}\cH_{BC}=0}$, ${\na_{A}d=0}$, see Table~\ref{TableDFT}. Eq.(\ref{deltacH0}) holds because the DFT-metric is constrained to be a symmetric $\ODD$ element~(\ref{defining}), see also  (\ref{twoCON}) below. This is    the reason  why  in the  variation of the action~(\ref{AV}) $\delta\cH_{AB}$  is contracted with a projected quantity, \textit{i.e.~}$(PG\brP)^{AB}-8\pi G(PT\brP)^{AB}$.  Eq.(\ref{AV}) is  then supposed to give the EDFEs, $G_{AB}=8\pi GT_{AB}$ (\ref{EDFE})~\cite{Angus:2018mep},  as the two variations, $\delta\cH_{AB}$ and $\delta d$, give the   projected part and the trace part separately,
\be
\ba{ll}
(PG\brP)_{AB}=8\pi G(PT\brP)_{AB}\,,
\qquad&\qquad
G_{A}{}^{A}=8\pi GT_{A}{}^{A}\,,
\ea
\label{PGtr}
\ee
which comprise the full EDFEs. While there is no issue on  the equation of motion of the DFT-dilaton, \textit{i.e.~}the  trace part in (\ref{PGtr}),  there might be some ambiguity on the DFT-metric variation especially around a non-Riemannian background. For example,  let us take one of the two maximally non-Riemannian, fully $\ODD$ symmetric vacua,  as  ${\cH_{AB}=\cJ_{AB}}$. Because  it does not allow any infinitesimal variation or moduli, ${\delta\cH_{AB}=0}$~\cite{Cho:2018alk},  the  induced variation of the action is null and therefore  it should not generate any nontrivial  Euler--Lagrange  equation of motion.  Nevertheless, in this case the `barred'  projector  vanishes automatically, ${\brP_{AB}=0}$,  and   the  projected part of the EDFEs in (\ref{PGtr})  is satisfied   rather trivially.    It appears that we have a slightly puzzling situation for the non-Riemannian background, ${\cH_{AB}=\cJ_{AB}}$: it allows no infinitesimal variation~$\delta\cH_{AB}=0$ and hence one may expect that the variation of the action should be trivial and there should be no nontrivial Euler--Lagrange equation of motion of the DFT-metric.  This is all true,  but nevertheless    the full EDFEs are  still valid!  (though in a trivial manner  as $\brP=0$).

Below, through  sections~\ref{SECvDFTm} and \ref{SECpG}, we shall  rigorously  revisit the variational principle of DFT around  a generic non-Riemannian background.  Basically,    we  are   questioning   whether it is really safe  from (\ref{deltacH0}) to put $\delta\cH_{AB}=2P_{(A}{}^{C}\brP_{B)}{}^{D}\cM_{CD}$ and read off the Euler--Lagrange equation of motion of the DFT-metric as if  $\cM_{CD}$ is a generic symmetric matrix. To answer this, we shall directly identify the truly independent degrees of freedom in the infinitesimal fluctuations of an  arbitrary  $(n,\brn)$ non-Riemannian DFT-metric, as  (\ref{indep}). We shall confirm that the full Einstein Double Field Equations   are still valid   for  non-Riemannian sectors,    either   trivially due to  projection properties    or nontrivially from the  genuine  variational principle.

\subsection{Variations of the DFT-metric  around a generic  $(n,\brn)$ background\label{SECvDFTm}}
Here  we shall identify the most general form of  the infinitesimal fluctuations around a generic $(n,\brn)$ DFT-metric~(\ref{cHFINAL}). The fluctuations   must  respect the defining properties of the DFT-metric~(\ref{defining}) and hence satisfy  
\be
\ba{ll}
\delta\cH_{AB}=\delta\cH_{BA}\,,\qquad&\qquad
\delta\cH_{A}{}^{B}\cH_{B}{}^{C}+
\cH_{A}{}^{B}\delta\cH_{B}{}^{C}=0\,.
\ea
\label{twoCON}
 \ee
It follows that $\delta\cH_{A}{}^{B}=-\cH_{A}{}^{C}\delta\cH_{C}{}^{D}\cH_{D}{}^{B}$, and hence equivalent (\ref{deltacH0}) holds. In particular,  $\delta\cH_{A}{}^{B}$ is traceless,
\be
\delta\cH_{A}{}^{A}=0\,.
\label{deltatrace}
\ee
That is to say, the trace of the DFT-metric, $\cH_{A}{}^{A}=2(n-\brn)$, is invariant under  continuous deformations.

Without loss of generality, like (\ref{DECOMP}), we put  
\be
\ba{lll}
\delta\cH_{AB}=\left(\ba{cc}1&0\\B&1\ea\right)\left(\ba{cr}\alpha~&~
\gamma\\
\gamma^{T}~&~ \beta
\ea\right)\left(\ba{cc}1&-B\\0&1\ea\right),\quad&\quad\alpha=\alpha^{T}\,,\quad&\quad\beta=\beta^{T}\,.
\ea
\label{deltahcHabg}
\ee
With this ansatz, the  former condition in (\ref{twoCON}) is met  and  the latter gives
\be
\ba{l}
\gamma Y_{i}(X^{i})^{T}-\gamma \brY_{\bri}(\brX^{\bri})^{T}+\alpha K+
Y_{i}(X^{i})^{T}\gamma-\brY_{\bri}(\brX^{\bri})^{T}\gamma+H\beta=0\,,\\
\beta Y_{i}(X^{i})^{T}-\beta \brY_{\bri}(\brX^{\bri})^{T}+\gamma^{T}K+
K\gamma+X^{i}(Y_{i})^{T}\beta-\brX^{\bri}(\brY_{\bri})^{T}\beta=0\,,\\
\gamma H+\alpha X^{i}(Y_{i})^{T}-\alpha\brX^{\bri}(\brY_{\bri})^{T}+Y_{i}(X^{i})^{T}\alpha- \brY_{\bri}(\brX^{\bri})^{T}\alpha+H\gamma^{T}=0\,.
\ea
\label{threecon}
\ee
We need to  solve these three constraints. For this, we utilize the completeness relation~(\ref{vcomp}), and decompose  each of $\{\alpha,\beta,\gamma\}$ into  mutually  orthogonal  pieces,
\be
\ba{lll}
\alpha^{\mu\nu}&=&H^{\mu}{}_{a}H^{\nu}{}_{b}\alpha^{ab}+
Y^{\mu}_{i}Y^{\nu}_{j}\alpha^{ij}+\brY^{\mu}_{\bri}\brY^{\nu}_{\brj}\alpha^{\bri\brj}+
2H^{(\mu}{}_{a}Y^{\nu)}_{i}\alpha^{ai}
+2H^{(\mu}{}_{a}\brY^{\nu)}_{\bri}\alpha^{a\bri}+
2Y^{(\mu}{}_{i}\brY^{\nu)}_{\bri}\alpha^{i\bri}\,,\\
\beta_{\mu\nu}&=&K_{\mu}{}^{a}K_{\nu}{}^{b}\beta_{ab}+
X_{\mu}^{i}X_{\nu}^{j}\beta_{ij}+\brX_{\mu}^{\bri}\brX_{\nu}^{\brj}\beta_{\bri\brj}+
2K_{(\mu}{}^{a}X_{\nu)}^{i}\beta_{ai}
+2K_{(\mu}{}^{a}\brX_{\nu)}^{\bri}\beta_{a\bri}+
2X_{(\mu}{}^{i}\brX_{\nu)}^{\bri}\beta_{i\bri}\,,\\
\gamma^{\mu}{}_{\nu}&=&H^{\mu}{}_{a}K_{\nu}{}^{b}\gamma^{a}{}_{b}
+
H^{\mu}{}_{a}X_{\nu}^{i}\gamma^{a}{}_{i}
+H^{\mu}{}_{a}\brX_{\nu}^{\bri}\gamma^{a}{}_{\bri}
+Y^{\mu}_{i}K_{\nu}{}^{a}\gamma^{i}{}_{a}+
Y^{\mu}_{i}X_{\nu}^{j}\gamma^{i}{}_{j}
+Y^{\mu}_{i}\brX_{\nu}^{\brj}\gamma^{i}{}_{\brj}\\
{}&{}&
+\brY^{\mu}_{\bri}K_{\nu}{}^{a}\gamma^{\bri}{}_{a}
+\brY^{\mu}_{\bri}X_{\nu}^{j}\gamma^{\bri}{}_{j}
+\brY^{\mu}_{\bri}\brX_{\nu}^{\brj}\gamma^{\bri}{}_{\brj}\,,
\ea
\label{abg}
\ee
where, since $\alpha,\beta$ are symmetric, 
\be
\ba{llllll}
\alpha^{ab}=\alpha^{ba}\,,\quad&\quad
\alpha^{ij}=\alpha^{ji}\,,\quad&\quad
\alpha^{\bri\brj}=\alpha^{\brj\bri}\,,\quad&\quad
\beta_{ab}=\beta_{ba}\,,\quad&\quad
\beta_{ij}=\beta_{ji}\,,\quad&\quad
\beta_{\bri\brj}=\beta_{\brj\bri}\,.
\ea
\ee
We remind the readers  that, using  the $(D-n-\brn)$-dimensional flat metric, $\eta_{ab}$, we  freely raise  or lower  the indices, $a,b$.  Now,  with the  decomposition~(\ref{abg}), it is straightforward to see that (\ref{threecon}) implies
\be
\ba{rrrr}
\alpha_{a}{}^{i}+\gamma^{i}{}_{ a}=0\,,\quad&\quad
\alpha_{a}{}^{\bri}-\gamma^{\bri}{}_{a}=0\,,\quad&\quad
\beta_{a i}+\gamma_{a i}=0\,,\quad&\quad
\beta_{a\bri}-\gamma_{a\bri}=0\,,\\
\alpha_{ab}+\beta_{ab}=0\,,\quad&\quad
\gamma_{ab}+\gamma_{ba}=0\,,\quad&\quad
\alpha^{ij}=0\,,\quad&\quad
\alpha^{\bri\brj}=0\,,\\
\beta_{ij}=0\,,\quad&\quad
\beta_{\bri\brj}=0\,,\quad&\quad
\gamma^{i}{}_{j}=0\,,\quad&\quad
\gamma^{\bri}{}_{\brj}=0\,.
\ea
\label{abgresult}
\ee
Thus, the independent degrees of  freedom for  the fluctuations  consist of
\be
\ba{ccccc}
\alpha_{(ab)}=-\beta_{(ab)}\,,\quad&\quad
\gamma_{[ab]}\,,\quad&\quad
\gamma^{a}{}_{i}=-\beta^{a}{}_{i}\,,\quad&\quad
\gamma^{a}{}_{\bri}=\beta^{a}{}_{\bri}\,,\quad&\quad
\gamma^{i}{}_{a}=-\alpha_{a}{}^{i}\,,\\
\gamma^{\bri}{}_{a}=\alpha_{a}{}^{\bri}\,,\quad&\quad
\alpha^{i\bri}\,,\quad&\quad
\beta_{j\brj}\,,\quad&\quad
\gamma^{i}{}_{\bri}\,,\quad&\quad
\gamma^{\brj}{}_{j}\,.
\ea
\label{indep}
\ee
In total,  as counted sequently  as
\be
\half(D-n-\brn)(D-n-\brn+1)+
\half(D-n-\brn)(D-n-\brn-1)+
2(D-n-\brn)(n+\brn)+4n\brn=D^{2}-(n-\brn)^{2}\,,
\label{idcH}
\ee
there are  $D^{2}-(n-\brn)^{2}$ number of degrees of freedom  which matches  precisely  the dimension of the underlying  coset~\cite{Berman:2019izh},
 \be
 \frac{\ODD}{\mathbf{O}(t+n,s+n)\times\mathbf{O}(s+\brn,t+\brn)}\,.
\label{COSET}
 \ee
Furthermore,  if we employ the DFT-vielbeins,\footnote{The only required property of the DFT-vielbeins  is  $\,V_{Ap}V_{B}{}^{p}+\brV_{A\brp}\brV_{B}{}^{\brp}=\cJ_{AB}$. See \cite{Cho:2018alk} for a related discussion.} $V_{Ap}, \brV_{A\brp}$,  the projected part of the EDFEs~(\ref{PGtr}) is equivalent to
\be
\left[(PG\brP)_{AB}-8\pi G(PT\brP)_{AB}\right]V^{A}{}_{p}\brV^{B}{}_{\brp}=0\,.
\label{vPG}
\ee
As the local Lorentz vector indices $p$ and $\brp$ run from one to ${D+n-\brn}$ and ${D-n+\brn}$ respectively,  there are in total $(D+n-\brn)\times(D-n+\brn)=D^{2}-(n-\brn)^{2}$ number of components in (\ref{vPG}) which coincides with the total number of independent fluctuations of the $(n,\brn)$ DFT-metric~(\ref{idcH}). As the number of the equations and the  fluctuations are the same, we may well expect that the former should be implied by the variational principle generated by the latter. Below,  we confirm this directly through explicit computation, without using  the DFT-vielbeins.

\subsection{Einstein Double Field Equations  still hold    for non-Riemannian  sectors \label{SECpG}}
Now, we proceed to organize the  variation of the action induced by  that of the $(n,\brn)$ DFT-metric~(\ref{AV}) in terms of the independent degrees of  freedom for the fluctuations~(\ref{indep}).

We apply the prescription~(\ref{gDCMP}) and write  a pair of circled  `$B$-field-free'  symmetric projectors,
\be
\ba{l}
\mP_{AB}=\mP_{BA}=(\cB^{-1})_{A}{}^{C}(\cB^{-1})_{B}{}^{D}P_{CD}
=
\frac{1}{2}\left(\ba{cc}H&
HK+2Y_{i}(X^{i})^{T}\\
KH+2X^{i}(Y_{i})^{T}&\quad K
\ea\right),\\
\mbrP_{AB}=\mbrP_{BA}=(\cB^{-1})_{A}{}^{C}(\cB^{-1})_{B}{}^{D}\brP_{CD}
=
\frac{1}{2}\left(\ba{cc}-H&
HK+2\brY_{\bri}(\brX^{\bri})^{T}\\
KH+2\brX^{\bri}(\brY_{\bri})^{T}&\quad -K
\ea\right),
\ea
\label{circled}
\ee
which satisfy  $\mP_{A}{}^{B}+\mbrP_{A}{}^{B}=\delta_{A}{}^{B}$\,, $\,\mP_{A}{}^{B}\mbrP_{B}{}^{C}=0$\,,  and    useful identities, 
\be
\ba{lll}
K_{\mu a}\mP^{\mu}{}_{A}=H^{\mu}{}_{a}\mP_{\mu A}\,,\qquad&\qquad
\brX_{\mu}^{\bri}\mP^{\mu}{}_{A}=0\,,\qquad&\qquad \brY^{\mu}_{\bri}\mP_{\mu A}=0\,,\\
K_{\mu a}\mbrP^{\mu}{}_{A}=-H^{\mu}{}_{a}\mbrP_{\mu A}\,,\qquad&\qquad
X_{\mu}^{i}\mbrP^{\mu}{}_{A}=0\,,\qquad&\qquad Y^{\mu}_{i}\mbrP_{\mu A}=0\,.
\ea
\label{uptouse}
\ee
We also introduce a  shorthand notation for the Einstein Double Field Equations,
\be
\ba{ll}
E_{AB}:=G_{AB}-8\pi GT_{AB}\,,\qquad&\qquad
\hE_{AB}:=(\cB^{-1})_{A}{}^{C}(\cB^{-1})_{B}{}^{D}E_{CD}\,.
\ea
\label{hE}
\ee
Hereafter,  hatted  quantities   contain generically the $\Hf$-flux, 
\be     
\Hf_{\lambda\mu\nu}=\partial_{\lambda}B_{\mu\nu}+\partial_{\mu}B_{\nu\lambda}
+\partial_{\nu}B_{\lambda\mu}\,,
\ee
but, like the circled ones,   there is no apparent bare $B$-field in them.

 It is  now straightforward to compute the variation in (\ref{AV}),
\be
\ba{l}
\delta\cH_{AB}(PE\brP)^{AB}\\
=\,2\gamma^{a}{}_{i}X_{\mu}^{i} (\mP\hE\mbrP)^{\mu}{}_{\nu}H^{\nu}{}_{a}
+2\gamma^{a}{}_{\bri}H^{\mu}{}_{a}(\mP\hE\mbrP)_{\mu}{}^{\nu}\brX_{\nu}^{\bri}
-2\gamma^{i}{}_{a}Y^{\mu}_{i}(\mP\hE\mbrP)_{\mu\nu}H^{\nu a}
+2\gamma^{\bri}{}_{a}H^{\mu}{}_{a}(\mP\hE\mbrP)_{\mu\nu}\brY^{\nu}_{\bri}\\
\quad\,+\alpha^{i\bri}Y^{\mu}_{i}(\mP\hE\mbrP)_{\mu\nu}\brY^{\nu}_{\bri}
+\gamma^{i}{}_{\bri}Y^{\mu}_{i}(\mP\hE\mbrP)_{\mu}{}^{\nu}\brX_{\nu}^{\bri}
+\gamma^{\bri}{}_{i}X_{\mu}^{i}(\mP\hE\mbrP)^{\mu}{}_{\nu}\brY^{\nu}_{\bri}
+\beta_{i\bri}X_{\mu}^{i}(\mP\hE\mbrP)^{\mu\nu}\brX_{\nu}^{\bri}\\
\quad\,+2\left(\alpha^{(ab)}-\gamma^{[ab]}\right)H^{\mu}{}_{a}(\mP\hE\mbrP)_{\mu\nu}H^{\nu}{}_{b}\,.
\ea
\ee
Each term is independent and thus, from the variational principle, should vanish individually on-shell,
\be
\ba{rrr}
X_{\mu}^{i} (\mP\hE\mbrP)^{\mu}{}_{\nu}H^{\nu}{}_{a}=0\,,\qquad&\qquad
H^{\mu}{}_{a}(\mP\hE\mbrP)_{\mu}{}^{\nu}\brX_{\nu}^{\bri}=0\,,\qquad&\qquad
Y^{\mu}_{i}(\mP\hE\mbrP)_{\mu\nu}H^{\nu}{}_{a}=0\,,\\
H^{\mu}{}_{a}(\mP\hE\mbrP)_{\mu\nu}\brY^{\nu}_{\bri}=0\,,\qquad&\qquad
Y^{\mu}_{i}(\mP\hE\mbrP)_{\mu\nu}\brY^{\nu}_{\bri}=0\,,\qquad&\qquad
Y^{\mu}_{i}(\mP\hE\mbrP)_{\mu}{}^{\nu}\brX_{\nu}^{\bri}=0\,,\\
X_{\mu}^{i}(\mP\hE\mbrP)^{\mu}{}_{\nu}\brY^{\nu}_{\bri}=0\,,\quad&\quad
X_{\mu}^{i}(\mP\hE\mbrP)^{\mu\nu}\brX_{\nu}^{\bri}=0\,,\quad&\quad
H^{\mu}{}_{a}(\mP\hE\mbrP)_{\mu\nu}H^{\nu}{}_{b}=0\,.
\ea
\label{nine}
\ee
In total,  as counted sequently as, 
\be
2(D-n-\brn)(n+\brn)+4n\brn+(D-n-\brn)^{2}=D^{2}-(n-\brn)^{2}\,,
\label{countEDFEs}
\ee
there is ${D^{2}-(n-\brn)^{2}}$ number of independent on-shell relations, or EDFEs, in  consistent with    (\ref{idcH}).

Up to the completeness relations~(\ref{COMP}), (\ref{vcomp}),  and the identities~(\ref{uptouse}), 
the first and the seventh in (\ref{nine}), 
the first and the eighth, 
the third and  the fifth,
 the third and the sixth,     
 the second and the last, 
the fourth and the last imply respectively,
\be
\ba{cc}
X_{\mu}^{i}(\mP\hE\mbrP)^{\mu}{}_{\nu}=0\,,\qquad\quad
X_{\mu}^{i}(\mP\hE\mbrP)^{\mu\nu}=0\,,\qquad&\quad
Y^{\mu}_{i}(\mP\hE\mbrP)_{\mu\nu}=0\,,\qquad\quad
Y^{\mu}_{i}(\mP\hE\mbrP)_{\mu}{}^{\nu}=0\,,\\
H^{\mu}{}_{a}(\mP\hE\mbrP)_{\mu}{}^{\nu}=K_{\mu a}
(\mP\hE\mbrP)^{\mu\nu}=0\,,\quad&\quad
H^{\mu}{}_{a}(\mP\hE\mbrP)_{\mu\nu}=K_{\mu a}(\mP\hE\mbrP)^{\mu}{}_{\nu}=0\,.
\ea
\ee
Finally, the first and the last, the second and the fifth, the third and the last, the fourth and the fifth give
\be
\ba{llll}
(\mP\hE\mbrP)^{\mu}{}_{\nu}=0\,,\qquad&\quad
(\mP\hE\mbrP)^{\mu\nu}=0\,,\qquad&\quad
(\mP\hE\mbrP)_{\mu\nu}=0\,,\qquad&\quad
(\mP\hE\mbrP)_{\mu}{}^{\nu}=0\,.
\ea
\label{EDFE2}
\ee
In this way,  all the components of  $(\mP\hE\mbrP)_{AB}$ vanish and  the full EDFEs  persist to be valid   universally for arbitrary $(n,\brn)$ backgrounds. \\

\noindent\textbf{Comment.} From (\ref{uptouse}),    off-shell relations hold among the components  of  the EDFEs,
\be{{
\ba{cc}
(\mP\hE\mbrP)_{\mu}{}^{\nu}=K_{\mu\rho}(\mP\hE\mbrP)^{\rho\nu}+X^{i}_{\mu}Y_{i}^{\rho}
(\mP\hE\mbrP)_{\rho}{}^{\nu}\,,\qquad&\quad
(\mP\hE\mbrP)^{\mu}{}_{\nu}=-(\mP\hE\mbrP)^{\mu\rho}K_{\rho\nu}+
(\mP\hE\mbrP)^{\mu}{}_{\rho}\brY_{\bri}^{\rho}\brX_{\nu}^{\bri}\,,\\
\multicolumn{2}{c}{\!\!\!\!\!(\mP\hE\mbrP)_{\mu\nu}=
-K_{\mu\rho}(\mP\hE\mbrP)^{\rho\sigma}K_{\sigma\nu}+
K_{\mu\rho}(\mP\hE\mbrP)^{\rho}{}_{\sigma}\brY_{\bri}^{\sigma}\brX_{\nu}^{\bri}
-X^{i}_{\mu}Y_{i}^{\rho}
(\mP\hE\mbrP)_{\rho}{}^{\sigma}K_{\sigma\nu}+
X^{i}_{\mu}Y_{i}^{\rho}(\mP\hE\mbrP)_{\rho\sigma}\brY_{\bri}^{\sigma}\brX_{\nu}^{\bri}\,,
}
\ea}}
\label{offshell}
\ee
such that the full EDFEs  are satisfied if
\be
\ba{lllll}
(\mP\hE\mbrP)^{\mu\nu}=0\,,\qquad&\quad
Y_{i}^{\mu}
(\mP\hE\mbrP)_{\mu}{}^{\nu}=0\,,\qquad&\quad
(\mP\hE\mbrP)^{\mu}{}_{\nu}\brY_{\bri}^{\nu}=0\,,\qquad&\quad
Y_{i}^{\mu}
(\mP\hE\mbrP)_{\mu\nu}\brY^{\nu}_{\bri}=0\,,\qquad&\quad\hE_{A}{}^{A}=0\,.
\ea
\label{EDFEconciseS}
\ee


\section{What if we keep $(n,\brn)$   fixed once and for all\,?\label{SECfix}}

As it is a outstandingly  hard problem  to construct  an action principle for  non-Riemannian gravity~(\textit{c.f.~}\cite{Bergshoeff:2016lwr,Hartong:2016yrf,Hansen:2018ofj}  for  recent proposals),   we may ask if the DFT action restricted to a fixed $(n,\brn)$ sector might serve as the desired target spacetime gravitational action, \textit{c.f.~}(\ref{Sofixed}).  In this section, seeking  for the answer to this question, we reanalyze the variational principle of DFT, crucially keeping  $(n,\brn)$ fixed. To our surprise, we  obtain  a subtle discrepancy with  the previous  section where the most general variations of the DFT-metric were  analyzed. We shall see that,  when the values of  $(n,\brn)$ are kept   fixed and $n\brn\neq0$, not  all the components of the EDFEs~(\ref{EDFEconciseS}) are implied by the variational principle.

\subsection{Variational principle  with fixed $(n,\brn)$}
We start with (\ref{AV}) which gives  the variation of the general DFT action induced  by the DFT-metric.   With fixed $(n,\brn)$, the variation of the DFT-metric therein  should comprise the variations of  the  $(n,\brn)$ component  fields:
\be
\!\!\delta\cH=\cB{\small{\left(\ba{cc}
\delta H&-H\delta B+
\delta\!\left[Y_{i}(X^{i})^{T}-\brY_{\bri}(\brX^{\bri})^{T}\right]\\
\delta B H+
\delta\!\left[X^{i}(Y_{i})^{T}-\brX^{\bri}(\brY_{\bri})^{T}\right]
&~~
\delta K+
\delta B\left[Y_{i}(X^{i})^{T}-\brY_{\bri}(\brX^{\bri})^{T}\right]
-\left[(X^{i}(Y_{i})^{T}-\brX^{\bri}(\brY_{\bri})^{T}\right]\delta B
\ea\right)\!\cB^{T}\,.}}
\label{deltacH}
\ee
Further, from their defining relations,   (\ref{HXX}),  (\ref{COMP}),   the variations of the   $(n,\brn)$ component  fields are not entirely independent. They must meet
\be
\ba{l}
\delta Y^{\mu}_{i}=-H^{\mu\rho}\delta K_{\rho\sigma}Y^{\sigma}_{i}-Y^{\mu}_{j}\delta X^{j}_{\rho}Y^{\rho}_{i}-\brY^{\mu}_{\brj}\delta\brX^{\brj}_{\rho}Y^{\rho}_{i}\,,\\
\delta \brY^{\mu}_{\bri}=-H^{\mu\rho}\delta K_{\rho\sigma}\brY^{\sigma}_{\bri}-Y^{\mu}_{j}\delta X^{j}_{\rho}\brY^{\rho}_{\bri}-\brY^{\mu}_{\brj}\delta\brX^{\brj}_{\rho}\brY^{\rho}_{\bri}\,,\\
\delta X_{\mu}^{i}=-K_{\mu\rho}\delta H^{\rho\sigma}X_{\sigma}^{i}-X_{\mu}^{j}\delta Y_{j}^{\rho}X_{\rho}^{i}-\brX_{\mu}^{\brj}\delta \brY_{\brj}^{\rho}X_{\rho}^{i}\,,\\
\delta\brX_{\mu}^{\bri}=-K_{\mu\rho}\delta H^{\rho\sigma}\brX_{\sigma}^{\bri}-X_{\mu}^{j}\delta Y_{j}^{\rho}\brX_{\rho}^{\bri}-\brX_{\mu}^{\brj}\delta \brY_{\brj}^{\rho}\brX_{\rho}^{\bri}\,,
\ea
\label{GIV1}
\ee
\be
\ba{l}
\delta H^{\mu\nu}=-H^{\mu\rho}\delta K_{\rho\sigma}H^{\sigma\nu}-2Y_{i}^{(\mu}H^{\nu)\rho}
\delta X^{i}_{\rho}-2\brY_{\bri}^{(\mu}H^{\nu)\rho}
\delta\brX^{\bri}_{\rho}\,,\\
\delta K_{\mu\nu}=-K_{\mu\rho}\delta H^{\rho\sigma}K_{\sigma\nu}-2\delta Y_{i}^{\rho}K_{\rho(\mu}X^{i}_{\nu)}-2\delta \brY_{\bri}^{\rho}K_{\rho(\mu}\brX^{\bri}_{\nu)}\,.
\ea
\label{GIV2}
\ee
From (\ref{KHroots}), we also note
\be
\ba{ll}
\delta K_{\mu\nu}=2K_{(\mu}{}^{a}\delta K_{\nu) a}\,,
\qquad&\qquad
\delta H^{\mu\nu}=2H^{(\mu}{}_{a}\delta H^{\nu) a}\,,
\ea
\label{deltaKH}
\ee
which imply in particular,
\be
\ba{ll}
\delta K_{\mu\nu}Y^{\mu}_{i}\brY^{\nu}_{\bri}=0\,,\qquad&\qquad
\delta H^{\mu\nu}X_{\mu}^{i}\brX_{\nu}^{\bri}=0\,.
\ea
\label{deltaKH2}
\ee
It is then evident  from (\ref{GIV1}), (\ref{GIV2}),  and (\ref{deltaKH}) that we have freedom to choose  either $\left\{\delta K_{\mu}{}^{a},\delta X^{i}_{\rho},\delta\brX^{\bri}_{\sigma}\right\}$ or  $\left\{\delta H^{\mu}{}_{a},\delta Y_{i}^{\rho},\delta\brY_{\bri}^{\sigma}\right\}$ as independent  variations. Each of them  has (formally)  $D^{2}$ number of degrees of freedom.

Now, we substitute  (\ref{deltacH})  into   (\ref{AV}), and utilize   (\ref{GIV1}), (\ref{GIV2}),  (\ref{deltaKH}), (\ref{deltaKH2}) to obtain 
\be
\ba{l}
\delta\dis{\int}\textstyle{\frac{1}{16\pi G}}\,e^{-2d}\So+\cL_{{{\rm{matter}}}}\\
=\dis{\int}
\textstyle{\frac{1}{4\pi G}}\,e^{-2d}\left[
2\delta K_{\nu a}K_{\mu}{}^{a}(\mP\hE\mbrP)^{(\mu\nu)}
+Y_{i}^{\rho}(\mP\hE\mbrP)_{\rho}{}^{\mu}\delta X^{i}_{\mu}
-\delta\brX^{\bri}_{\mu}(\mP\hE\mbrP)^{\mu}{}_{\rho}\brY^{\rho}_{\bri}
-\delta B_{\mu\nu}(\mP\hE\mbrP)^{\mu\nu}
\right]\\
=\dis{\int}
\textstyle{\frac{1}{4\pi G}}\,e^{-2d}\left[
2\delta H^{\nu a} H^{\mu}{}_{a}(\mP\hE\mbrP)_{(\mu\nu)}
+X^{i}_{\rho}(\mP\hE\mbrP)^{\rho}{}_{\mu}\delta Y_{i}^{\mu}
-\delta\brY_{\bri}^{\mu}(\mP\hE\mbrP)_{\mu}{}^{\rho}\brX_{\rho}^{\bri}
-\delta B_{\mu\nu}(\mP\hE\mbrP)^{\mu\nu}
\right].
\ea
\label{apEL}
\ee
The variational  principle implies either from the second line of (\ref{apEL}),
\be
\ba{llll}
K_{\mu\rho}(\mP\hE\mbrP)^{\rho\nu}+(\mP\hE\mbrP)^{\nu\rho}K_{\rho\mu}= 0\,,\quad&\quad\!
Y_{i}^{\rho}(\mP\hE\mbrP)_{\rho}{}^{\mu}= 0\,,\quad&\quad\!
(\mP\hE\mbrP)^{\mu}{}_{\rho}\brY^{\rho}_{\bri}= 0\,,\quad&\quad\!
(\mP\hE\mbrP)^{[\mu\nu]}= 0\,,
\ea
\label{KPSsym}
\ee
or alternatively from the third  line of (\ref{apEL}),
\be
\ba{llll}
H^{\mu\rho}(\mP\hE\mbrP)_{\rho\nu}+(\mP\hE\mbrP)_{\nu\rho}H^{\rho\mu}= 0\,,\quad&\quad\!
X^{i}_{\rho}(\mP\hE\mbrP)^{\rho}{}_{\mu}= 0\,,\quad&\quad\!
(\mP\hE\mbrP)_{\mu}{}^{\rho}\brX_{\rho}^{\bri}= 0\,,\quad&\quad\!
(\mP\hE\mbrP)^{[\mu\nu]}= 0\,.
\ea
\label{HPSsym}
\ee
Although (\ref{KPSsym}) and (\ref{HPSsym}) appear seemingly different,  they  are ---as should be---  equivalent.  In fact, they are both equivalent  to
\be
\ba{llll}
(\mP\hE\mbrP)^{\mu\nu}=0\,,\quad&\quad
(\mP\hE\mbrP)^{\mu}{}_{\nu}=0\,,\quad&\quad
(\mP\hE\mbrP)_{\mu}{}^{\nu}=0\,,\quad&\quad
(\mP\hE\mbrP)_{\mu\nu}=X_{\mu}^{i}Y_{i}^{\rho}(\mP\hE\mbrP)_{\rho\sigma}\brY^{\sigma}_{\bri}\brX^{\bri}_{\nu}\,,
\ea
\label{nbrnEDFE}
\ee
which are,   from (\ref{offshell}), further  equivalent  to more concise ones,
\be
\ba{llll}
(\mP\hE\mbrP)^{\mu\nu}=0\,,\quad&\qquad
(\mP\hE\mbrP)^{\mu}{}_{\nu}\brY^{\nu}_{\bri}=0\,,\quad&\qquad
Y^{\mu}_{i}(\mP\hE\mbrP)_{\mu}{}^{\nu}=0\,.
\ea
\label{nbrnEDFEconcise}
\ee
Appendix~\ref{SECProof} carries  our proof.  

The surprise  which  is manifest in (\ref{nbrnEDFE}) is that,  when $n\brn\neq0$   the variational principle with fixed $(n,\brn)$   does not imply  the full EDFEs~(\ref{EDFEconciseS}):  it  does not constrain  $\,Y_{i}^{\rho}(\mP\hE\mbrP)_{\rho\sigma}\brY^{\sigma}_{\bri}$. However,  as we have shown in the previous section,   within the DFT frame   they should vanish on-shell, $\,Y_{i}^{\rho}(\mP\hE\mbrP)_{\rho\sigma}\brY^{\sigma}_{\bri}=0$, and the full EDFEs should  hold.   We shall continue to discuss  and conclude    in   the  final section~\ref{SECconclusion}.

\subsection{Difference between  keeping  $(n,\brn)$ fixed   or not}
In order to understand  the discrepancy in the resulting Euler--Lagrangian equations,  (\ref{EDFE2}) \textit{vs.} (\ref{nbrnEDFE}),  here we  investigate   how    the infinitesimal  variations of the component fields of the  $(n,\brn)$ DFT-metric~(\ref{deltacH}), 
\be
\big\{\delta H^{\mu\nu}\,,\,\delta K_{\rho\sigma}\,,\,\delta X^{i}_{\mu}\,,\,\delta Y^{\nu}_{j}\,,\,\delta \brX^{\bri}_{\rho}\,,\,\delta\brY^{\sigma}_{\brj}\,,\,\delta B_{\mu\nu}\big\}\,,
\label{nnfield}
\ee
contribute   actually  to the $\alpha,\beta,\gamma$ variables defined in the generic variation of the DFT-metric~(\ref{deltahcHabg}),
\be
\!\!\!{\small{\left(\!\ba{cr}\alpha~&~
\gamma\\
\gamma^{T}~&~ \beta
\ea\!\right)
=\left(\!\!\ba{cc}
\delta H&-H\delta B+
\delta\!\left[Y_{i}(X^{i})^{T}-\brY_{\bri}(\brX^{\bri})^{T}\right]\\
\delta B H+
\delta\!\left[X^{i}(Y_{i})^{T}-\brX^{\bri}(\brY_{\bri})^{T}\right]
&
\delta K+
\delta B\left[Y_{i}(X^{i})^{T}-\brY_{\bri}(\brX^{\bri})^{T}\right]
-\left[(X^{i}(Y_{i})^{T}-\brX^{\bri}(\brY_{\bri})^{T}\right]\delta B
\ea\!\!\right).}}
\ee
With (\ref{abg}),  one can identify the contributions  thoroughly:
\be
\ba{ll}
\alpha_{ab}=-H^{\mu}{}_{a}H^{\nu}{}_{b}\delta K_{\mu\nu}=-2\delta K_{\rho(a}H^{\rho}{}_{b)}\,,\qquad
&\qquad  
\beta_{ab}=-\alpha_{ab}=-2K_{\rho(a}\delta H^{\rho}{}_{b)}=-K_{\mu a}K_{\nu b}\delta H^{\mu\nu}\,,\\
\alpha^{a i}=-H^{\rho a}\delta X^{i}_{\rho}\,,\qquad&\qquad
\beta_{a i}=-K_{\rho a}\delta Y^{\rho}_{i}+H^{\rho}{}_{a}\delta  B_{\rho\sigma}Y^{\sigma}_{i}\,,\\
\alpha^{a\bri}=-H^{\rho a}\delta\brX^{\bri}_{\rho}\,,\qquad&\qquad
\beta_{a\bri}=-K_{\rho a}\delta\brY^{\rho}_{\bri}-
H^{\rho}{}_{a}\delta B_{\rho\sigma}\brY^{\sigma}_{\bri}\,, \\
\alpha^{ij}=0\,,\qquad&\qquad \beta_{ij}=0\,,\\
\alpha^{\bri\brj}=0\,,
\qquad&\qquad \beta_{\bri\brj}=0\,,\\
\!\!\!\begin{boxed}{\,
{\alpha^{i\bri}=0\,,}\,}
\end{boxed}
 \qquad&\qquad\beta_{i\bri}=-2Y_{i}^{\rho}\delta B_{\rho\sigma}\brY^{\sigma}_{\bri}\,,
 \ea
\label{alphavanish}
 \ee
 and
 \be
 \ba{ll}
\gamma^{a}{}_{i}=K_{\rho}{}^{a}\delta Y_{i}^{\rho}-H^{\rho a}\delta B_{\rho\sigma}Y^{\sigma}_{i}=-\beta^{a}{}_{i}\,,\qquad&\qquad
\gamma^{a}{}_{\bri}=-K_{\rho}{}^{a}\delta\brY_{\bri}^{\rho}-H^{\rho a}\delta B_{\rho\sigma}\brY^{\sigma}_{\bri}=\beta^{a}{}_{\bri}\,,\\
\gamma^{i}{}_{a}=-X^{i}_{\rho}\delta H^{\rho\sigma} K_{\sigma a}=-\alpha_{a}{}^{i}\,,\qquad&\qquad
\gamma^{\bri}{}_{a}=\brX^{\bri}_{\rho}\delta H^{\rho\sigma} K_{\sigma a}=\alpha_{a}{}^{\bri}\,,\\
\gamma^{i}{}_{j}=0\,,\qquad&\qquad
\gamma^{\bri}{}_{\brj}=0\,,\\
\gamma^{i}{}_{\bri}=-X^{i}_{\rho}\delta\brY_{\bri}^{\rho}\,,\qquad&\qquad
\gamma^{\bri}{}_{i}=\brX^{\bri}_{\rho}\delta Y_{i}^{\rho}\,,\\
\gamma_{ab}=-\gamma_{ba}=-H^{\rho}{}_{a}H^{\sigma}{}_{b}\delta B_{\rho\sigma}\,.\qquad&\qquad {}
\ea
\ee
This  is consistent with the  general result of (\ref{abgresult}). However, one surprise is that $\alpha^{i\bri}$ must be trivial when the $(n,\brn)$ component fields~(\ref{nnfield}) are varied while keeping $(n,\brn)$  fixed.

To identify  the significance of the $\alpha^{i\bri}$ parameter,  we  focus on the  induced transformation of   $H^{\mu\nu}$,  
\be
\ba{lll}
H^{\mu\nu}~&~\longrightarrow~&~H^{\prime\mu\nu}\simeq
H^{\mu\nu}+2Y^{(\mu}_{i}\brY^{\nu)}_{\bri}\alpha^{i\bri}\,.
\ea
\ee
Geometrically the deformation of $2Y^{(\mu}_{i}\brY^{\nu)}_{\bri}\alpha^{i\bri}$ is `orthogonal' to $H^{\mu\nu}$, and thus  we expect  it  should reduce the kernel of $H^{\mu\nu}$. To verify this explicitly, we solve for the eigenvectors of $H^{\prime\mu\nu}$ with  zero eigenvalue,
\be
H^{\prime\mu\nu}\cX_{\nu}=0\,.
\label{kernelcX}
\ee
Without loss of generality, utilizing   the completeness relation, $K_{\mu a}H^{\nu a}+X_{\mu}^{i}Y^{\nu}_{i}+\brX_{\mu}^{\bri}\brY^{\nu}_{\bri}=\delta_{\mu}{}^{\nu}$,   we decompose  the zero-eigenvector,
\be
\cX_{\nu}=K_{\nu a}c^{a}+X^{i}_{\nu}c_{i}+\brX^{\bri}_{\nu}\brc_{\bri}\,,
\ee
substitute  this ansatz  into (\ref{kernelcX}), and  acquire  the  conditions the coefficients should satisfy,  
\be
\ba{lll}
c^{a}=0\,,\quad&\quad
\alpha^{i\bri}c_{i}=0\,, \quad&\quad\alpha^{i\bri}\brc_{\bri}=0\,.
\ea
\ee
This shows  that there are  in total $(n-\mbox{rank\,}[\alpha^{i\bri}])+(\brn-\mbox{rank\,}[\alpha^{i\bri}])=n+\brn-2\times \mbox{rank\,}[\alpha^{i\bri}]$ number of zero-eigenvectors.  Moreover, from the invariance, ${\delta\cH_{A}{}^{A}=0}$~(\ref{deltatrace}), we note that the deformation  by the $\alpha^{i\bri}$ parameter actually changes the type of the `non-Riemannianity' as
\be
\ba{lll}
(n,\brn)~&~\longrightarrow~&~\left(n-\mbox{rank\,}[\alpha^{i\bri}]\,,\,\brn-\mbox{rank\,}[\alpha^{i\bri}]\right).
\ea
\label{changenbrn}
\ee
This   essentially explains why $\alpha^{i\bri}$  vanishes in (\ref{alphavanish}) where   the $(n,\brn)$ component  field variables are varied with fixed  values of   $(n,\brn)$, or fixed `non-Riemannianity'.  It is  intriguing to note that the deformation  makes the DFT-metric  always \textit{less} non-Riemannian.\footnote{In a way, on the space of full DFT geometries, the $(0,0)$ Riemannian geometry corresponds to an open set as $\det(H^{\mu\nu})\neq 0$,  while the genuine non-Riemannian geometries  form  a closed set, $\det(H^{\mu\nu})= 0$.  Infinitesimally, it is impossible to leave an open set but    possible to leave a  closed set.   }\\



\subsection{Non-Riemannian  differential geometry  as bookkeeping device\label{SECtool}}
\begin{center}\textit{This subsection is the last one before Conclusion, and is somewhat out of  context. At first reading, \\  readers may glimpse  (\ref{Sofixed}) in comparison with (\ref{SoRiemann}), and  skip to the final  section~\ref{SECconclusion}. }\end{center}

 While the  various $(n,\brn)$ non-Riemannian geometries are universally well described by DFT through  $\ODD$ covariant tensors ---as summarized in Table~\ref{TableDFT}---  it may be desirable  in practical computations  to break   the  manifest $\ODD$ symmetry spontaneously  by fixing the section, $\tpartial^{\mu}\equiv0$, and  dismantle the $\ODD$ covariant tensors  or   curvatures into smaller  \textit{modules} which should be still covariant under  undoubled   ordinary diffeomorphisms, $B$-field gauge symmetry, and $\GL(n)\times\GL(\brn)$ local rotations.  We remind the readers that in the case of the $(0,0)$ Riemannian sector, the $\ODD$ singlet  DFT scalar curvature  reduces to  four  modules (\textit{c.f.~}\cite{Andriot:2013xca,Blair:2014zba,Lee:2016qwn}):
\be
\left.\So\right|_{(0,0)~{\rm{Riemannian}}}=
R_{g}
-\textstyle{\frac{1}{12}}\Hf^{\lambda\mu\nu}\Hf_{\lambda\mu\nu}
+4\Box\phi-4\partial^{\mu}\phi\partial_{\mu}\phi\,.
\label{SoRiemann}
\ee
Here in this last subsection,    we propose  an undoubled non-Riemannian   differential tool kit, such as  covariant derivative and curvature,  for an arbitrary $(n,\brn)$ sector.  It descends  from the DFT geometry, or the  so-called  ``semi-covariant formalism"~\cite{Jeon:2011cn}, and generalizes the standard  Riemannian geometry underlying (\ref{SoRiemann}) in a consistent manner.  It breaks the manifest $\ODD$ symmetry spontaneously,  but preserves  the  ordinary diffeomorphisms, $B$-field gauge symmetry,  and the  $\GL(n)\times\GL(\brn)$ local symmetries as desired.  In particular, it enables us to extend  the Riemannian  expression of (\ref{SoRiemann})  in a way `continuously'   to the generic $(n,\brn)$ non-Riemannian case,
\be
\ba{rcl}
\So\Big|_{(n,\brn)~\rm{fixed}}\!&\!\!=&\!
R-\textstyle{\frac{1}{12}}H^{\lambda\rho}H^{\mu\sigma}H^{\nu\tau}\Hf_{\lambda\mu\nu}\Hf_{\rho\sigma\tau}
-\Hf_{\lambda\mu\nu}H^{\lambda\rho}\!\left(Y^{\mu}_{i}\MD^{\nu}X^{i}_{\rho}
-\brY^{\mu}_{\bri}\MD^{\nu}\brX^{\bri}_{\rho}\right)\\
{}&{}&+4K_{\mu\nu}\!
\left(\MD^{\mu}\MD^{\nu}d-\MD^{\mu}d\,\MD^{\nu}d\right).
\ea
\label{Sofixed}
\ee

We commence our  explanation. First of all,   $\MD^{\mu}$ is  our  proposed  `upper-indexed'  covariant derivative:
\be
\MD^{\mu}=H^{\mu\rho}\partial_{\rho}+\Omega^{\mu}+\Upsilon^{\mu}+\brUpsilon^{\mu}\,,
\label{MD}
\ee
which preserves   both  the undoubled diffeomorphisms~(\ref{oLie}) and  the   $\GL(n)\times\GL(\brn)$  local symmetries~(\ref{deltaw}) as is  equipped   with proper connections:  for   undoubled ordinary diffeomorphisms, 
\be
\ba{lll}
\Omega^{\mu\nu}{}_{\lambda}&=&-\half\partial_{\lambda}H^{\mu\nu}
- H^{\rho[\mu}\partial_{\rho}H^{\nu]\sigma}K_{\sigma\lambda}
-H^{\rho[\mu}\partial_{\rho}Y_{i}^{\nu]}X^{i}_{\lambda}-H^{\rho[\mu}\partial_{\rho}\brY_{\bri}^{\nu]}\brX^{\bri}_{\lambda}\\
{}&{}&
+\left(2
H^{\rho[\mu}Y_{i}^{\nu]}\partial_{[\tau}X^{i}_{\rho]}
-2H^{\rho[\mu}\brY_{\bri}^{\nu]}\partial_{[\tau}\brX^{\bri}_{\rho]}
\right)\!\left(Y_{j}^{\tau}X_{\lambda}^{j}-\brY_{\brj}^{\tau}\brX_{\lambda}^{\brj}\right),
\ea
\label{Omega}
\ee
and   for  $\GL(n)\times\GL(\brn)$  rotations,
\be
\ba{ll}
\Upsilon^{\mu\, i}{}_{j}
=-2H^{\mu\rho}Y^{\sigma}_{j}\partial_{[\rho}X^{i}_{\sigma]}
\,,\qquad&\qquad
\brUpsilon^{\mu\,\bri}{}_{\brj}
=-2H^{\mu\rho}\brY^{\sigma}_{\brj}\partial_{[\rho}\brX^{\bri}_{\sigma]}\,.
\ea
\label{Upsilon}
\ee
We also denote  a diffeomorphism-only preserving  covariant derivative by 
\be
\fD^{\mu}=H^{\mu\rho}\partial_{\rho}+\Omega^{\mu}\,,
\label{deffD}
\ee
and write    for  (\ref{MD}) and (\ref{Upsilon}), 
\be
\ba{lll}
\MD^{\mu}
=\fD^{\mu}+\Upsilon^{\mu}+\brUpsilon^{\mu}\,,\qquad&\quad
\Upsilon^{\mu\, i}{}_{j}=X^{i}_{\rho}\fD^{\mu}Y_{j}^{\rho}
=-Y_{j}^{\rho}\fD^{\mu}X^{i}_{\rho}\,,\qquad&\quad
\brUpsilon^{\mu\,\bri}{}_{\brj}
=\brX^{\bri}_{\rho}\fD^{\mu}\brY_{\brj}^{\rho}=-\brY_{\brj}^{\rho}
\fD^{\mu}\brX^{\bri}_{\rho}\,.
\ea
\label{MDUU}
\ee
Taking care of both   spacetime  and $\GL(n)\times\GL(\brn)$ indices, 
$\MD^{\mu}$   acts   on general   tensor densities   in a standard manner:
\be
\ba{lll}
\MD^{\lambda}T^{\mu i\bri}{}_{\nu j\brj}&=&H^{\lambda\rho}\partial_{\rho}T^{\mu i\bri}{}_{\nu j\brj}-\omegaT\Omega^{\lambda\rho}{}{}_{\rho}T^{\mu i\bri}{}_{\nu j\brj}
+\Omega^{\lambda\mu}{}_{\rho}T^{\rho i\bri}{}_{\nu j\brj}
-\Omega^{\lambda\rho}{}_{\nu}T^{\mu i\bri}{}_{\rho j\brj}\\
{}&{}&
+\Upsilon^{\lambda i}{}_{k}T^{\mu k\bri}{}_{\nu j\brj}
+\brUpsilon^{\lambda\bri}{}_{\brk}T^{\mu i\brk}{}_{\nu j\brj}
-\Upsilon^{\lambda k}{}_{j}T^{\mu i\bri}{}_{\nu k\brj}
-\brUpsilon^{\lambda\brk}{}_{\brj}T^{\mu i\bri}{}_{\nu j\brk}\,.
\ea
\ee
On the other hand,     $\fD^{\mu}$ cares only the spacetime indices and ignores any  $\GL(n)\times\GL(\brn)$ indices,
\be
\fD^{\lambda}T^{\mu i\bri}{}_{\nu j\brj}=H^{\lambda\rho}\partial_{\rho}T^{\mu i\bri}{}_{\nu j\brj}-\omegaT\Omega^{\lambda\rho}{}{}_{\rho}T^{\mu i\bri}{}_{\nu j\brj}
+\Omega^{\lambda\mu}{}_{\rho}T^{\rho i\bri}{}_{\nu j\brj}
-\Omega^{\lambda\rho}{}_{\nu}T^{\mu i\bri}{}_{\rho j\brj}\,.
\ee 
For example, we have explicitly
\be
\ba{l}
\MD^{\mu}X^{i}_{\nu}=H^{\mu\rho}\partial_{\rho}X^{i}_{\nu}-X^{i}_{\rho}\Omega^{\mu\rho}{}_{\nu}+\Upsilon^{\mu i}{}_{j}X^{j}_{\nu}=H^{\mu\rho}(KH)_{\nu}{}^{\sigma}\partial_{[\rho}X^{i}_{\sigma]}\,,\\
\MD^{\mu}\brX^{\bri}_{\nu}
=H^{\mu\rho}\partial_{\rho}\brX^{\bri}_{\nu}-\brX^{\bri}_{\rho}\Omega^{\mu\rho}{}_{\nu}+\brUpsilon^{\mu\bri}{}_{\brj}\brX^{\brj}_{\nu}=H^{\mu\rho}(KH)_{\nu}{}^{\sigma}\partial_{[\rho}\brX^{\bri}_{\sigma]}\,.
\ea
\label{MDX}
\ee
It is instructive to see that the far right  resulting  expressions in (\ref{MDX}) are clearly covariant under both diffeomorphisms and $\GL(n)\times\GL(\brn)$ local rotations, as the $\rho$, $\sigma$ indices therein are skew-symmetrized and also contracted with $H^{\mu\rho}$, $(KH)_{\nu}{}^{\sigma}$.  However,  without the $\GL(n)\times\GL(\brn)$ connections, we note
\be
\fD^{\mu}X^{i}_{\nu}
=H^{\mu\rho}\partial_{\rho}X^{i}_{\nu}-\Omega^{\mu\rho}{}_{\nu}X^{i}_{\rho}
=H^{\mu\rho}\!\left[(KH)_{\nu}{}^{\sigma}+2X_{\nu}^{j}Y_{j}^{\sigma}\right]\partial_{[\rho}X^{i}_{\sigma]}\,,
\ee
and this breaks the  $\GL(n)\times\GL(\brn)$ local symmetry.

Further, for the DFT-dilaton we should have
\be
\MD^{\mu}d=\fD^{\mu}d=-\half e^{2d\,}\fD^{\mu}\!\left(e^{-2d}\right)=H^{\mu\rho}\partial_{\rho}d+\half\Omega^{\mu\rho}{}_{\rho}\,,
\ee
where we have explicitly 
\be
\Omega^{\mu\rho}{}_{\rho}=H^{\mu\nu}\!\left(\half H^{\rho\sigma}\partial_{\nu}K_{\rho\sigma}+ Y_{i}^{\rho}\partial_{\rho}X^{i}_{\nu}+\brY_{\bri}^{\rho}\partial_{\rho}\brX^{\bri}_{\nu}\right)
=-\half K_{\rho\sigma} \partial^{\mu}H^{\rho\sigma}
+K_{\rho\sigma}\partial^{\sigma}H^{\mu\rho}-\partial_{\rho}H^{\mu\rho}\,.
\ee
Because $H^{\mu\nu}$ and $K_{\rho\sigma}$ are generically degenerate, the conventional  relation~(\ref{Riemannian})  between the DFT-dilaton, $d$,  and  the  string dilaton, $\phi$, cannot hold. We stick to use the DFT-dilaton all the way.\footnote{We tend to believe that the conventional string dilaton, $\phi$,  is an artifact of the $(0,0)$ Riemannian geometry and the DFT-dilaton, $d$,  is more fundamental as being an  $\ODD$ singlet.}

The connections do the job as they transform properly  under the diffeomorphisms~(\ref{oLie}), (\ref{xMV}) and  the $\GL(n)\times\GL(\brn)$ local rotations~(\ref{deltaw}),
\be
\ba{ll}
\delta_{\xi}\Omega^{\mu\nu}{}_{\lambda}=\cL_{\xi}\Omega^{\mu\nu}{}_{\lambda}+H^{\mu\rho}\partial_{\rho}\partial_{\lambda}\xi^{\nu}\,,\qquad&\qquad
\deltaw\Omega^{\mu\nu}{}_{\lambda}=0\,,\\
\delta_{\xi}\Upsilon^{\mu i}{}_{j}=\cL_{\xi}\Upsilon^{\mu i}{}_{j}\,,\qquad&\qquad
\deltaw\Upsilon^{\mu i}{}_{j}=\Upsilon^{\mu k}w_{k}{}^{i}-w_{j}{}^{k}\Upsilon^{\mu i}{}_{k}-H^{\mu\rho}\partial_{\rho}w_{j}{}^{i}\,,\\
\delta_{\xi}\brUpsilon^{\mu \bri}{}_{\brj}=\cL_{\xi}\brUpsilon^{\mu \bri}{}_{\brj}\,,\qquad&\qquad
\deltaw\brUpsilon^{\mu \bri}{}_{\brj}=\brUpsilon^{\mu \brk}\brw_{\brk}{}^{\bri}-\brw_{\brj}{}^{\brk}\brUpsilon^{\mu \bri}{}_{\brk}-H^{\mu\rho}\partial_{\rho}\brw_{\brj}{}^{\bri}\,.
\ea
\ee
In particular, $X^{i}_{\mu}\Omega^{\mu\nu}{}_{\lambda}$, 
$\brX^{\bri}_{\mu}\Omega^{\mu\nu}{}_{\lambda}$, and $H^{\rho[\lambda}\Omega^{\mu]\nu}{}_{\rho}$ are  covariant tensors which might be viewed as    ``torsions".

Finally,  we define an upper-indexed Ricci curvature, 
\be
R^{\mu\nu}:=H^{\mu\rho}\partial_{\rho}\Omega^{\sigma\nu}{}_{\sigma}
-H^{\sigma\rho}\partial_{\rho}\Omega^{\mu\nu}{}_{\sigma}
+\Omega^{\mu\nu}{}_{\rho}\Omega^{\sigma\rho}{}_{\sigma}-
\Omega^{\sigma\mu}{}_{\rho}\Omega^{\rho\nu}{}_{\sigma}
+2\left(Y_{i}^{\sigma}\MD^{\mu} X^{i}_{\rho} 
+\brY_{\bri}^{\sigma}\MD^{\mu}\brX^{\bri}_{\rho}\right)\Omega^{\rho\nu}{}_{\sigma}\,,
\label{Ricci}
\ee
which is diffeomorphism  and $\GL(n)\times\GL(\brn)$ covariant, as it comes from  the following commutator relation that is clearly also covariant,
\be
\left[\MD^{\mu},\MD^{\nu}\right]
T_{\nu}+4\left(Y_{i}^{\sigma}\MD^{\mu} X^{i}_{\rho} 
+\brY_{\bri}^{\sigma}\MD^{\mu}\brX^{\bri}_{\rho}\right)H^{\rho\nu}
\partial_{[\sigma}T_{\nu]}
+2\left(
Y_{i}^{\nu}\MD^{\mu} X^{i}_{\rho} 
+\brY_{\bri}^{\nu}\MD^{\mu}\brX^{\bri}_{\rho}\right)\MD^{\rho}T_{\nu}
=-R^{\mu\nu}T_{\nu}\,.
\ee
A scalar curvature follows naturally, 
\be
R:=K_{\mu\nu}R^{\mu\nu}\,,
\label{scalarR1}
\ee
which  debuted  in (\ref{Sofixed}).\\

Our covariant derivative  is  ``compatible" with the $(n,\brn)$ component fields in a generalized fashion:
\be
\ba{ll}
\MD^{\lambda}H^{\mu\nu}+2Y_{i}^{(\mu}H^{\nu)\rho}\MD^{\lambda}X^{i}_{\rho}+
2\brY_{\bri}^{(\mu}H^{\nu)\rho}\MD^{\lambda}\brX^{\bri}_{\rho}=0\,,\qquad&\qquad 
Y_{i}^{\rho}\MD^{\mu}X^{j}_{\rho}=0\,,
\qquad\brY_{\bri}^{\rho}\MD^{\mu}\brX^{\brj}_{\rho}=0\,, \\
\MD^{\lambda}K_{\mu\nu}+2X^{i}_{(\mu}K_{\nu)\rho}\MD^{\lambda}Y^{\rho}_{i}
+2\brX^{\bri}_{(\mu}K_{\nu)\rho}\MD^{\lambda}\brY^{\rho}_{\bri}=0\,,\qquad&\quad\MD^{\lambda}\delta_{\mu}{}^{\nu}=0\,,\qquad
\MD^{\lambda}\delta_{i}{}^{j}=0\,,\qquad \MD^{\lambda}\delta_{\bri}{}^{\brj}=0
\,.
\ea
\label{COMP1}
\ee
Another characteristic is that, if we add   one more  torsion  linear in  the $\Hf$-flux to the $\Omega$-connection,
\be
\ba{ll}
\hOmega^{\mu\nu}{}_{\lambda}:=\Omega^{\mu\nu}{}_{\lambda}
+\half H^{\mu\rho}H^{\nu\sigma}\Hf_{\rho\sigma\tau}
\!\left(Y_{j}^{\tau}X_{\lambda}^{j}-\brY_{\brj}^{\tau}\brX_{\lambda}^{\brj}\right),\qquad&\quad \hfD^{\mu}:=H^{\mu\rho}\partial_{\rho}+\hOmega^{\mu}\,,
\ea
\label{hOmega}
\ee
the hatted  new connection becomes  Milne-shift  covariant  as well,  in  the sense of  (\ref{MS}),  (\ref{xMV}),  (\ref{Mcov2}),
\be
\ba{ll}
\deltaM\hOmega^{\mu\nu}{}_{\lambda}=-\half\deltaM B_{\lambda\rho}\hH^{\mu\nu\rho}\,,\qquad&\qquad
\deltaM\hH^{\lambda\mu\nu}=0\,,
\ea
\label{MhOhH}
\ee
where $\hH^{\lambda\mu\nu}$ is a diffeomorphism covariant,  $\GL(n)\times\GL(\brn)$ invariant, and  Milne-shift invariant $\Hf$-flux,
\be
\hH^{\lambda\mu\nu}=\hH^{[\lambda\mu\nu]}:=
H^{\lambda\rho}H^{\mu\sigma}H^{\nu\tau}H_{\rho\sigma\tau}
+6H^{\rho[\lambda}Y^{\mu}_{i}\MD^{\nu]}X^{i}_{\rho}
-6H^{\rho[\lambda}\brY^{\mu}_{\bri}\MD^{\nu]}\brX^{\bri}_{\rho}\,.
\label{hHdef}
\ee
The $\GL(n)\times\GL(\brn)$ connections~(\ref{MDUU})  are inert  to the addition of the  $\Hf$-flux-valued-torsion~(\ref{hOmega}) as
\be
\ba{ll}
\Upsilon^{\mu\, i}{}_{j}=X^{i}_{\rho}\fD^{\mu}Y_{j}^{\rho}
=X^{i}_{\rho}\hfD^{\mu}Y_{j}^{\rho}=\widehat{\Upsilon}{}^{\mu\,i}{}_{j}\,,\qquad&\quad
\brUpsilon^{\mu\,\bri}{}_{\brj}
=\brX^{\bri}_{\rho}\fD^{\mu}\brY_{\brj}^{\rho}=\brX^{\bri}_{\rho}
\hfD^{\mu}\brY_{\brj}^{\rho}=\widehat{\brUpsilon}{}^{\mu\,\bri}{}_{\brj}\,,
\ea
\label{inert}
\ee
while they  transform  under the Mine-shift as $\deltaM\Upsilon^{\mu i}{}_{j}=
-2H^{\rho\sigma}V_{\rho j}\MD^{\mu}X^{i}_{\sigma\,}$, 
$\,\deltaM\brUpsilon^{\mu\bri}{}_{\brj}=-2H^{\rho\sigma}\brV_{\rho\brj}\MD^{\mu}\brX^{\bri}_{\sigma\,}$.

After all,  in terms of   a hatted covariant derivative,
\be
\hMD^{\mu}:=H^{\mu\rho}\partial_{\rho}
+\hOmega^{\mu}+\widehat{\Upsilon}{}^{\mu}+\widehat{\brUpsilon}{}^{\mu}\,,
\label{hMDdef}
\ee
we can   dismantle the DFT curvatures into   {\darkblue{a $\Hf$-flux-free (circled) term}}  and  evidently  $\Hf$-flux-valued  ones: 
\be
\!\!\ba{rcl}
\So\!\!&\!\!\!=&\!\!\! 
{\darkblue{\mSo}}
-\textstyle{\frac{1}{12}}H^{\lambda\rho}H^{\mu\sigma}H^{\nu\tau}\Hf_{\lambda\mu\nu}\Hf_{\rho\sigma\tau}
-\Hf_{\lambda\mu\nu}H^{\lambda\rho}
\left(Y^{\mu}_{i}\hMD^{\nu}X^{i}_{\rho}-\brY^{\mu}_{\bri}\hMD^{\nu}\brX^{\bri}_{\rho}\right),\\
Y^{\mu}_{i}(\mP\hS\mbrP)_{\mu}{}^{\nu}\!\!&\!\!\!=&\!\!\!
{\darkblue{ Y^{\mu}_{i}(\mP\mS\mbrP)_{\mu}{}^{\nu}}}
+Y^{\mu}_{i}\!\left[\Hf_{\mu\rho\sigma}\!\left(\brY_{\bri}^{[\rho}\hMD^{\nu]}\brX^{\bri}_{\lambda}-\half
Y^{\rho}_{j}\hMD^{\nu}X^{j}_{\lambda}\right)\!H^{\lambda\sigma}+\quarter H^{\nu\sigma}e^{2d}\hMD^{\rho}\left(e^{-2d}\Hf_{\rho\sigma\mu}\right)
\right],\\
(\mP\hS\mbrP)^{\mu}{}_{\nu}\brY_{\bri}^{\nu}\!\!&\!\!\!=&\!\!\! 
{\darkblue{(\mP\mS\mbrP)^{\mu}{}_{\nu}\brY_{\bri}^{\nu} }}
+\left[\Hf_{\rho\sigma\nu}\!\left(Y_{i}^{[\rho}\hMD^{\mu]}X^{i}_{\lambda}-\half
\brY^{\rho}_{\brj}\hMD^{\mu}\brX^{\brj}_{\lambda}\right)\!H^{\lambda\sigma}
+\quarter H^{\mu\sigma}e^{2d}\hMD^{\rho}\left(e^{-2d}\Hf_{\rho\sigma\nu}\right)
\right]\!\brY^{\nu}_{\bri}\,,\\
Y^{\mu}_{i}(\mP\hS\mbrP)_{\mu\nu}\brY_{\bri}^{\nu}\!\!&\!\!\!=&\!\!\!\! 
{\darkblue{Y^{\mu}_{i}(\mP\mS\mbrP)_{\mu\nu}\brY_{\bri}^{\nu} }}
+ \half Y^{\mu}_{i}\brY_{\bri}^{\nu}\Big[e^{2d}\hMD^{\rho}\left(e^{-2d}\Hf_{\rho\mu\nu}\right)+
 \half H^{\alpha\beta}H^{\gamma\delta}\Hf_{\mu\alpha\gamma}\Hf_{\nu\beta\delta}\Big]\,,\\
(\mP\hS\mbrP)^{\mu\nu}\!\!&\!\!\!=&\!\!\!
{\darkblue{(\mP\mS\mbrP)^{\mu\nu}}}
-\textstyle{\frac{1}{8}}e^{2d}\partial_{\lambda}(e^{-2d}\hH^{\lambda\mu\nu})
+\textstyle{\frac{1}{16}}H^{\mu\rho}H^{\nu\sigma}H^{\alpha\beta}H^{\gamma\delta}
\Hf_{\rho\alpha\gamma}\Hf_{\sigma\beta\delta}\\
{}&{}&~+\textstyle{\frac{3}{8}}\Big[
H^{\mu\rho}\!\left(H^{\lambda[\nu}Y_{i}^{\sigma}\hMD^{\tau]}X^{i}_{\lambda}
-H^{\lambda[\nu}\brY_{\bri}^{\sigma}\hMD^{\tau]}\brX^{\bri}_{\lambda}\right)
~+~(\mu~\leftrightarrow~\nu)~
\Big]\Hf_{\rho\sigma\tau}\,,
\label{SEPARATE}
\ea
\ee
where, as it  should be obvious from our  notation,   we set $\hS_{AB}:=(\cB^{-1})_{A}{}^{C}(\cB^{-1})_{B}{}^{D}S_{CD}$,   and    the  circled quantities are all $\Hf$-flux free: from  Table~\ref{TableDFT} or  \cite{Jeon:2011cn,Angus:2018mep},
\be
\mS_{AB}=2\partial_{A}\partial_{B}d-e^{2d\,}\partial_{C}\left(e^{-2d\,}\mGamma_{(AB)}{}^{C}\right)+
\half\mGamma_{ACD}\mGamma_{B}{}^{CD}-\half\mGamma_{CDA}\mGamma^{CD}{}_{B}\,,
\label{mSAB}
\ee
and, with (\ref{circled}),
\be
\ba{lrl}
\mGamma_{CAB}&:=&2(\mP\partial_{C}\mP\mbrP)_{[AB]}
+2({{\mbrP}_{[A}{}^{D}{\mbrP}_{B]}{}^{E}}-{\mP_{[A}{}^{D}\mP_{B]}{}^{E}})\partial_{D}\mP_{EC}\\
{}&{}&-4\left(\textstyle{\frac{1}{\mbrP_{G}{}^{G}-1}}\mbrP_{C[A}\mbrP_{B]}{}^{D}+\textstyle{\frac{1}{\mP_{G}{}^{G}-1}}\mP_{C[A}\mP_{B]}{}^{D}\right)\left(\partial_{D}d
+(\mP\partial^{E}\mP\mbrP)_{[ED]}\right).
\ea
\label{mGamma}
\ee
While we organize the $\Hf$-flux-valued  parts in terms of the hatted covariant derivative, like (\ref{inert}),  we have
\be
\ba{llll}
\hMD^{\mu}X^{i}_{\nu}=\MD^{\mu}X^{i}_{\nu}\,, \qquad&\quad\hMD^{\mu}\brX^{\bri}_{\nu}=\MD^{\mu}\brX^{\bri}_{\nu}\,,\qquad&\quad \hMD^{\mu}d=\MD^{\mu}d\,,\qquad&\quad\hMD^{\mu}\hMD^{\nu}d=\MD^{\mu}\MD^{\nu}d\,.
\ea
\ee
The only nontrivial distinction lies in
\be
\hMD^{\rho}\left(e^{-2d}\Hf_{\rho\mu\nu}\right)=
\MD^{\rho}\left(e^{-2d}\Hf_{\rho\mu\nu}\right)+
H^{\rho\alpha}H^{\sigma\beta}H_{\rho\sigma\tau}\!\left(
H_{\alpha\beta[\mu}X^{i}_{\nu]}Y_{i}^{\tau}-
H_{\alpha\beta[\mu}\brX^{\bri}_{\nu]}\brY_{\bri}^{\tau}\right).
\ee
Since $e^{-2d}\hH^{\lambda\mu\nu}$ carries a unit  weight, its contraction with the ordinary derivative, $\partial_{\lambda}(e^{-2d}\hH^{\lambda\mu\nu})$, is  also by itself   diffeomorphism covariant.  In this way,  every single term in (\ref{SEPARATE})   is  symmetric under both undoubled  diffeomorphisms and $\GL(n)\times\GL(\brn)$  local rotations. On the other hand, as we have singled out  the $\Hf$-flux-valued  terms from the  $\Hf$-flux-free parts,  each individual   term is   not necessarily  Milne-shift covariant.

As advertised in (\ref{Sofixed}), we may further dismantle $\mSo$  as well as $(\mP\mS\mbrP)^{\mu\nu}$ into more elementary modules:
\be
\ba{rcl}
\mSo\!\!&\!\!=&\!\!\! 
R+4K_{\mu\nu}\left(\MD^{\mu}\MD^{\nu}d-\MD^{\mu}d\,\MD^{\nu}d\right),\\
(\mP\mS\mbrP)^{\mu\nu}\!\!&\!\!=&\!\!\! 
-\quarter R^{(\mu\nu)}
-\quarter\big(Y^{\mu}_{i}\MD^{\rho}X^{i}_{\sigma}-
\brY^{\mu}_{\bri}\MD^{\rho}\brX^{\bri}_{\sigma}\big)
\big(Y^{\nu}_{j}\MD^{\sigma}X^{j}_{\rho}-
\brY^{\nu}_{\brj}\MD^{\sigma}\brX^{\brj}_{\rho}\big)-\half\MD^{(\mu}\MD^{\nu)}d\,.
\ea
\label{furtherdismantle}
\ee
From  (\ref{EDFEconciseS}), vanishing of  all the  five quantities in (\ref{SEPARATE}) characterizes the   $(n,\brn)$  \textit{vacuum} geometry of DFT.\\

\noindent\textbf{Comment 1.}  It is worth while to note  
\be
e^{-2d}K_{\mu\nu}\left(\MD^{\mu}\MD^{\nu}d-2\MD^{\mu}d\,\MD^{\nu}d\right)=\partial_{\mu}\left(e^{-2d}\MD^{\mu}d\right)\,,
\ee
and rewrite the `kinetic term' of the DFT-dilaton in (\ref{Sofixed}),
\be
4e^{-2d}K_{\mu\nu}\left(\MD^{\mu}\MD^{\nu}d-\MD^{\mu}d\,\MD^{\nu}d\right)=
4e^{-2d}K_{\mu\nu}\MD^{\mu}d\,\MD^{\nu}d+4\partial_{\mu}\left(e^{-2d}\MD^{\mu}d\right).
\label{Sofixed2}
\ee
\noindent\textbf{Comment 2.}    Especially when $n+\brn=D$, \textit{i.e.~}in the maximally non-Riemannian cases,  all the quantities like $H^{\mu\nu},K_{\rho\sigma},\Omega^{\lambda\mu}{}_{\nu},
\hH^{\lambda\mu\nu},\MD^{\mu}d, R^{\mu\nu},\So,  (\mP\hS\mbrP)^{\mu\nu}$ are trivial except  the term of interest, 
$Y^{\mu}_{i}(\mP\hS\mbrP)_{\mu\nu}\brY^{\mu}_{\bri}$.

\noindent\textbf{Comment 3.}    Restricted  to the $(0,0)$ Riemannian case,  we have 
$K_{\mu\nu}=g_{\mu\nu\,}$, $\,H^{\mu\nu}=g^{\mu\nu}$,  $\,K_{\mu\rho}H^{\rho\mu}=\delta_{\mu}{}^{\nu}$,   
and  the vectors, $\{X_{\mu}^{i},\brX_{\nu}^{\bri},Y^{\rho}_{j},\brY^{\sigma}_{\brj}\}$,  are trivially  absent. Both the $\Omega$ and $\hOmega$ connections~(\ref{Omega},\ref{hOmega})   coincide with nothing but    the standard   Christoffel symbols with one index raised by the Riemannian  metric,
\be
\hOmega^{\mu\nu}{}_{\lambda}\equiv\Omega^{\mu\nu}{}_{\lambda}\equiv g^{\mu\rho}{\left\{{}^{\,\,\,\nu\,}_{\rho~\lambda}\right\}}\,.
\ee
Consequently,  the  proposed covariant derivative~(\ref{deffD}) and   Ricci curvature~(\ref{Ricci}) reduce to the standard covariant derivative and  Ricci curvature in  Riemannian geometry,
\be
\ba{ll}
\fD^{\mu}\equiv g^{\mu\nu}\trd_{\nu}=g^{\mu\nu}\big(\partial_{\nu}+
\left\{{}^{\,\,~\cdot\,}_{\nu~~\cdot}\right\}\big)\,,\qquad&\qquad
R^{\mu\nu}\equiv  g^{\mu\rho}g^{\nu\sigma}R^{\trd}_{\rho\sigma}\,.
\ea
\ee

\noindent\textbf{Comment 4.}  Besides $(\mP\mS\mbrP)^{\mu\nu}$, we have not been able to  dismantle other circled  $\Hf$-flux-free DFT Ricci curvatures which carry at least one lower index. In addition to  $\fD^{\mu}=H^{\mu\rho}\partial_{\rho}+\Omega^{\mu}$, separate  type of covariant derivatives  containing   $Y^{\mu}_{i}\partial_{\mu}$ or $\brY^{\mu}_{\bri}\partial_{\mu}$ might help, \textit{c.f.~}(\ref{YDT}).

\noindent\textbf{Comment 5.}  Appendix~\ref{SECDerivation}  sketches how we have arrived  at the above proposal   of  the  non-Riemannian differential tool kit   starting  from the semi-covariant  formalism of DFT.  In any case, our proposal is meant to provide  a bookkeeping device to expound the EDFEs into smaller modules and to  single out the $\Hf$-fluxes. The actual computation  of the variations of the action, even with  $(n,\brn)$ fixed,   are  still powered by the semi-covariant formalism,  specifically (\ref{POWER}).\\

\newpage

\section{Conclusion\label{SECconclusion}}
The very gravitational theory  string theory   predicts  may  be  the  Double Field Theory with non-Riemannian surprises, rather than General Relativity based on Riemannian geometry.     The underlying mathematical  structure of DFT unifies supergravity  with  various non-Riemannian gravities including   (stringy) Newton--Cartan geometry,   ultra-relativistic Carroll geometry,  and non-relativistic Gomis--Ooguri string theory.   The non-Riemannian geometries of DFT can be  classified by two non-negative integers, $(n,\brn)$~\cite{Morand:2017fnv}.

We have analyzed with care the variational principle.  We  have shown that the most general infinitesimal variations of an arbitrary  $(n,\brn)$ DFT-metric have  $D^{2}-(n-\brn)^{2}$ number of degrees of freedom, which  matches with the dimension of the underlying  coset~\cite{Berman:2019izh}, $ \frac{\ODD}{\mathbf{O}(t+n,s+n)\times\mathbf{O}(s+\brn,t+\brn)}$  (\ref{COSET}).  Through action principle,  these variations  imply   the full Einstein Double Field Equations~(\ref{countEDFEs}), (\ref{EDFE2}).  However,   $n\brn$ number of them  change  the value of $(n,\brn)$,  \textit{i.e.~}the type of non-Riemannianity~(\ref{changenbrn}).  Consequently, if we keep $(n,\brn)$   fixed once and for all, the variational principle gets  restricted   and fails to reproduce  the full EDFEs:  the specific part, $Y^{\mu}_{i}(PE\brP)_{\mu\nu}\brY^{\nu}_{\bri}$,  does  not have to  vanish on-shell~(\ref{nbrnEDFE}).\footnote{As can be  seen from (\ref{mSAB}), $\,Y_{i}^{\rho}(\mP\hE\mbrP)_{\rho\sigma}\brY^{\sigma}_{\bri}$    contains    a second order  derivative  of the DFT-dilaton  along the  $Y_{i}^{\mu}$ and $\brY_{\nu}^{\bri}$ directions, \textit{i.e.}~$\,Y^{\mu}_{i}\brY^{\nu}_{\bri}\partial_{\mu}\partial_{\nu}d\,$.    }

The   EDFEs are supposed to arise as the  string worldsheet beta-functions~\cite{Berman:2007xn,Copland:2011wx}.  For the doubled-yet-gauged string action~(\ref{stringaction}) upon an arbitrarily  chosen    $(n,\brn)$ background, the  $(n,\brn)$-changing variations of the DFT-metric  would  correspond to    marginal deformations.  We must stress  that   these deformations  could not   be realized  by merely varying   the  background component fields  with   fixed $(n,\brn)$~(\ref{alphavanish}), \textit{c.f.~}\cite{Gomis:2019zyu,Gallegos:2019icg,Bergshoeff:2019pij}.    Nevertheless,   it is natural  to expect that  $n\brn$ number of   $Y^{\mu}_{i}(\mP\hE\mbrP)_{\mu\nu}\brY^{\nu}_{\bri}$ arise  as the corresponding  beta-functions  too.   That is to say, at least for  $n\brn\neq0$, the quantum consistency with the worldsheet string theory seems to forbid us to fix $(n,\brn)$ rigidly.   We conclude   that the various non-Riemannian gravities should be identified  as different solution sectors of Double Field Theory rather than viewed as  independent theories.  Quantum consistency of the non-Riemannian geometries calls for  thorough   investigation, which  may enlarge the scope of the string  theory  landscape far beyond Riemann.

\section*{Acknowledgements}
We  would like to thank   David Berman, Chris Blair, and  Kevin Morand for helpful discussions.  Parts of  computations are assisted by a computer algebra system, \textit{Cadabra}~\cite{Peeters:2007wn}. This work  was  supported by  the National Research Foundation of Korea   through  the Grant,  NRF-2016R1D1A1B01015196.


\appendix

\begin{center}
{\bf\Large{APPENDIX}}
\end{center}
\section{Proof on the equivalence among (\ref{KPSsym}), (\ref{HPSsym}),  (\ref{nbrnEDFE}), and (\ref{nbrnEDFEconcise})\label{SECProof}}

Taking $\left\{\delta K_{\mu}{}^{a},\delta X^{i}_{\rho},\delta\brX^{\bri}_{\sigma}\right\}$  as independent  variations,  from the second line of (\ref{apEL}),  the variational  principle implies  (\ref{KPSsym})  which we enumerate here:
\begin{eqnarray}
K_{\mu\rho}(\mP\hE\mbrP)^{\rho\nu}+(\mP\hE\mbrP)^{\nu\rho}K_{\rho\mu}= 0\,,\label{KPS}\\
Y_{i}^{\rho}(\mP\hE\mbrP)_{\rho}{}^{\mu}= 0\,,\label{YPS}\\
(\mP\hE\mbrP)^{\mu}{}_{\rho}\brY^{\rho}_{\bri}= 0\,,\label{PSbrY}\\
(\mP\hE\mbrP)^{[\mu\nu]}= 0\,.\label{PSsym}
\end{eqnarray}
Alternatively taking  $\left\{\delta H^{\mu}{}_{a},\delta Y_{i}^{\rho},\delta\brY_{\bri}^{\sigma}\right\}$ as independent  variations, we acquire  from the third  line of (\ref{apEL}),
\begin{eqnarray}
H^{\mu\rho}(\mP\hE\mbrP)_{\rho\nu}+(\mP\hE\mbrP)_{\nu\rho}H^{\rho\mu}= 0\,,\label{HPS}\\
X^{i}_{\rho}(\mP\hE\mbrP)^{\rho}{}_{\mu}= 0\,,\label{XPS}\\
(\mP\hE\mbrP)_{\mu}{}^{\rho}\brX_{\rho}^{\bri}= 0\,,\label{PSbrX}\\
(\mP\hE\mbrP)^{[\mu\nu]}= 0\,.\label{PSsym2}
\end{eqnarray}
Henceforth   we show that  Eqs.(\ref{KPS},\ref{YPS},\ref{PSbrY},\ref{PSsym}) and 
Eqs.(\ref{HPS},\ref{XPS},\ref{PSbrX},\ref{PSsym2})  are all equivalent to (\ref{nbrnEDFE}) as well as to (\ref{nbrnEDFEconcise}). The equivalence between (\ref{nbrnEDFE}) and (\ref{nbrnEDFEconcise}) should be obvious  from the off-shell relation~(\ref{offshell}), and  therefore  we  focus on (\ref{nbrnEDFE}) which we recall for quick reference:
\be
\ba{llll}
(\mP\hE\mbrP)^{\mu\nu}=0\,,\quad&\quad
(\mP\hE\mbrP)^{\mu}{}_{\nu}=0\,,\quad&\quad
(\mP\hE\mbrP)_{\mu}{}^{\nu}=0\,,\quad&\quad
(\mP\hE\mbrP)_{\mu\nu}=X_{\mu}^{i}Y_{i}^{\rho}(\mP\hE\mbrP)_{\rho\sigma}\brY^{\sigma}_{\bri}\brX^{\bri}_{\nu}\,.
\ea
\label{nbrnEDFE2}
\ee

\textit{Proof.}\\
It is manifest that  (\ref{nbrnEDFE2}) implies both Eqs.(\ref{KPS},\ref{YPS},\ref{PSbrY},\ref{PSsym}) and 
Eqs.(\ref{HPS},\ref{XPS},\ref{PSbrX},\ref{PSsym2}). Thus, we only need to  show the converse.  Eq.(\ref{PSsym}) and Eq.(\ref{PSsym2})  are common and give,  combined with (\ref{uptouse}),
\be
\ba{ll}
X^{i}_{\mu}(\mP\hE\mbrP)^{\mu\nu}=0\,,\qquad&\qquad
(\mP\hE\mbrP)^{\mu\nu}\brX^{\bri}_{\nu}=0\,.
\ea
\label{manipC}
\ee
With these in mind we first focus on the former set of equations~(\ref{KPS},\ref{YPS},\ref{PSbrY},\ref{PSsym}), of which  the first and the last  imply
\be
\ba{ll}
K_{\mu\rho}(\mP\hE\mbrP)^{\rho\nu}= 0\,,\qquad&\qquad(\mP\hE\mbrP)^{\nu\rho}K_{\rho\mu}= 0\,.
\ea
\label{manip0}
\ee
Consequently,  with the completeness relation~(\ref{COMP}),  the identities from (\ref{uptouse}), and  (\ref{YPS}),  we note
\be
(\mP\hE\mbrP)_{\mu}{}^{\nu}=
(KH)_{\mu}{}^{\rho}(\mP\hE\mbrP)_{\rho}{}^{\nu}+X_{\mu}^{i}Y_{i}^{\rho}(\mP\hE\mbrP)_{\rho}{}^{\nu}
=K_{\mu\rho}(\mP\hE\mbrP)^{\rho\nu}+X_{\mu}^{i}Y_{i}^{\rho}(\mP\hE\mbrP)_{\rho}{}^{\nu}= 0\,.
\label{manip1}
\ee
Similarly  we get with (\ref{PSbrY}),
\be
(\mP\hE\mbrP)^{\mu}{}_{\nu}=
(\mP\hE\mbrP)^{\mu}{}_{\rho}(HK)^{\rho}_{\nu}+
(\mP\hE\mbrP)^{\mu}{}_{\rho}\brY^{\rho}_{\bri}\brX^{\bri}_{\nu}=
-(\mP\hE\mbrP)^{\mu\rho}K_{\rho\nu}+
(\mP\hE\mbrP)^{\mu}{}_{\rho}\brY^{\rho}_{\bri}\brX^{\bri}_{\nu}=0\,,
\label{manip2}
\ee
and with (\ref{uptouse}), (\ref{PSsym}),
\be
(\mP\hE\mbrP)^{\mu\nu}=(HK)^{\mu}{}_{\rho}(\mP\hE\mbrP)^{\rho\nu}= H^{\mu\rho}(\mP\hE\mbrP)_{\rho}{}^{\nu}=0\,.
\label{manip3}
\ee
It follows that
\be
\ba{l}
\!\!\!\!(\mP\hE\mbrP)_{\mu\nu}=\left[(KH)_{\mu}{}^{\rho}+X^{i}_{\mu}Y_{i}^{\rho}\right](\mP\hE\mbrP)_{\rho\sigma}\left[(HK)^{\sigma}{}_{\nu}+\brY_{\bri}^{\sigma}\brX^{\bri}_{\nu}\right]\\
{}\qquad\quad\,\, =
-K_{\mu\rho}(\mP\hE\mbrP)^{\rho\sigma}K_{\sigma\nu}+
K_{\mu\rho}(\mP\hE\mbrP)^{\rho}{}_{\sigma}\brY_{\bri}^{\sigma}\brX^{\bri}_{\nu}-
X^{i}_{\mu}Y_{i}^{\rho}(\mP\hE\mbrP)_{\rho}{}^{\sigma}K_{\sigma\nu}+
X^{i}_{\mu}Y_{i}^{\rho}(\mP\hE\mbrP)_{\rho\sigma}\brY_{\bri}^{\sigma}\brX^{\bri}_{\nu}\\
{}\qquad\quad\,\,=
X^{i}_{\mu}Y_{i}^{\rho}(\mP\hE\mbrP)_{\rho\sigma}\brY_{\bri}^{\sigma}\brX^{\bri}_{\nu}\,.
\ea
\label{manip4}
\ee
Thus, Eqs.(\ref{KPS},\ref{YPS},\ref{PSbrY},\ref{PSsym}) are equivalent to (\ref{nbrnEDFE2}).

We now turn to the latter  set of equations~(\ref{HPS},\ref{XPS},\ref{PSbrX},\ref{PSsym2}).   In a parallel manner to  
(\ref{manip1}), (\ref{manip2}),  we  note from (\ref{XPS}), (\ref{PSbrX}),
\be
\ba{l}
(\mP\hE\mbrP)_{\nu}{}^{\mu}
= (\mP\hE\mbrP)_{\nu}{}^{\rho}(KH)_{\rho}{}^{\mu}+ (\mP\hE\mbrP)_{\nu}{}^{\rho}\brX^{\bri}_{\rho}\brY^{\mu}_{\bri} 
=-
(\mP\hE\mbrP)_{\nu\rho}H^{\rho\mu}\,,\\
(\mP\hE\mbrP)^{\mu}{}_{\nu}
=(HK)^{\mu}{}_{\rho}(\mP\hE\mbrP)^{\rho}{}_{\nu}+Y^{\mu}_{i}X^{i}_{\rho}(\mP\hE\mbrP)^{\rho}{}_{\nu}
=
H^{\mu\rho}(\mP\hE\mbrP)_{\rho\nu}\,,
\ea
\ee
which imply  with (\ref{HPS}),
\be
(\mP\hE\mbrP)_{\mu}{}^{\nu}=(\mP\hE\mbrP)^{\nu}{}_{\mu}\,,
\ee
and hence in particular,
\be
\ba{llll}
Y^{\mu}_{i}(\mP\hE\mbrP)_{\mu}{}^{\nu}=0\,,\qquad&\qquad
(\mP\hE\mbrP)_{\mu}{}^{\nu}\brX^{\bri}_{\nu}=0\,,\qquad&\qquad
X^{i}_{\mu}(\mP\hE\mbrP)^{\mu}{}_{\nu}=0\,,\qquad&\qquad
(\mP\hE\mbrP)^{\mu}{}_{\nu}\brY^{\nu}_{\bri}=0\,.
\ea
\label{manip5}
\ee
Then  with (\ref{PSsym2})  and from
\be
\!(\mP\hE\mbrP)_{\mu}{}^{\nu}=
(KH)_{\mu}{}^{\rho}(\mP\hE\mbrP)_{\rho}{}^{\nu}
=K_{\mu\rho}(\mP\hE\mbrP)^{\rho\nu}=(\mP\hE\mbrP)^{\nu\rho}K_{\rho\mu}
=-(\mP\hE\mbrP)^{\nu}{}_{\rho}(HK)^{\rho}{}_{\mu}=-(\mP\hE\mbrP)^{\nu}{}_{\mu}\,,
\ee
we note 
\be
\ba{ll}
(\mP\hE\mbrP)_{\mu}{}^{\nu}=0\,,\qquad&\qquad
(\mP\hE\mbrP)^{\mu}{}_{\nu}=0\,.
\ea
\ee
It follows then,  with  (\ref{PSsym2}), (\ref{manipC}),
\be
(\mP\hE\mbrP)^{\mu\nu}=(HK)^{\mu}{}_{\rho}(\mP\hE\mbrP)^{\rho\nu}
= H^{\mu\rho}(\mP\hE\mbrP)_{\rho}{}^{\nu}=0\,.
\label{PSbrPup1}
\ee
Finally, as in (\ref{manip4}), we have
\be
(\mP\hE\mbrP)_{\mu\nu}=K_{\mu\rho}(\mP\hE\mbrP)^{\rho}{}_{\nu}+
X^{i}_{\mu}Y_{i}^{\rho}(\mP\hE\mbrP)_{\rho\sigma}\big[(HK)^{\sigma}{}_{\nu}+\brY^{\sigma}_{\bri}\brX_{\nu}^{\bri\,}\big]= X^{i}_{\mu}Y_{i}^{\rho}(\mP\hE\mbrP)_{\rho\sigma}\brY^{\sigma}_{\bri}\brX_{\nu}^{\bri}
\,.
\label{lastproof}
\ee
Thus, Eqs.(\ref{HPS},\ref{XPS},\ref{PSbrX},\ref{PSsym2}) are also equivalent to (\ref{nbrnEDFE2}), and this completes our  \textit{Proof}.



\section{Derivation of  the non-Riemannian differential tool kit  from DFT\label{SECDerivation} }
The  non-Riemannian differential geometry  we have proposed in section~\ref{SECtool}, in particular  the hatted $\hOmega$ connection~(\ref{hOmega}),   descends    from the known covariant derivatives in  the  DFT semi-covariant formalism~\cite{Jeon:2011cn}: specifically,\footnote{While $\na_{C}V_{D}$ itself  is not covariant, the projected ones in  (\ref{compCOV}) are covariant, and hence the name,  `semi-covariant formalism'.}    
\be
\ba{ll}
P_{A}{}^{C}\brP_{B}{}^{D}\na_{C}V_{D}\,,\qquad&\qquad
\brP_{A}{}^{C}P_{B}{}^{D}\na_{C}V_{D}\,.
\ea
\label{compCOV}
\ee
In order to convert these into undoubled ordinary  covariant  quantities ---or to get rid of the bare $B$-field in them---  we multiply    $\cB^{-1}$ as in (\ref{gDCMP}) and write 
\be
\ba{ll}
\!\!(\cB^{-1}P)^{AC}(\cB^{-1}\brP)^{BD}\na_{C}V_{D}
=\mP^{AC}\mbrP^{BD}\hat{\na}_{C}\mV_{D}\,,\qquad&\quad
(\cB^{-1}\brP)^{AC}(\cB^{-1}P)^{BD}\na_{C}V_{D}
=\mbrP^{AC}\mP^{BD}\hat{\na}_{C}\mV_{D}\,.
\ea
\label{cBcovD}
\ee
Here we set
\be
\hat{\na}_{A}\mV_{B}:=(\cB^{-1})_{A}{}^{C}(\cB^{-1})_{B}{}^{D}\na_{C}V_{D}=
\partial_{A}\mV_{B}+\hGamma_{ABC}\mV^{C}\,,
\ee
and $\hGamma_{ABC}$ is a naturally induced ---or `twisted'~\cite{Cho:2015lha}, \textit{c.f.~}\cite{Berman:2013cli}--- new connection,\footnote{Note  $\cB_{A}{}^{B}\partial_{B}=\partial_{A}$ as $\tpartial^{\mu}\equiv0$.}
\be
\ba{lrl}
\hGamma_{CAB}&:=&(\cB^{-1})_{C}{}^{D}(\cB^{-1})_{A}{}^{E}
(\cB^{-1})_{B}{}^{F}\Gamma_{DEF}+\partial_{C}\cB_{AB}\\
{}&=&\mGamma_{CAB}
+(\mbrP_{C}{}^{\rho}\mP_{A}{}^{\sigma}\mP_{B}{}^{\tau}+\mP_{C}{}^{\rho}\mbrP_{A}{}^{\sigma}\mbrP_{B}{}^{\tau})\Hf_{\rho\sigma\tau}
+({\mcP+\mbrcP})_{CAB}{}^{DEF}\partial_{D}\mcB_{EF}\,.
\ea
\label{Gammat2}
\ee
The  very last term on  the second line involves certain six-indexed projectors formed by $\mP_{A}{}^{B},\mbrP_{A}{}^{B}$~(\textit{c.f.} Eq.(17) of \cite{Jeon:2011cn} and Eq.(2.26) of \cite{Angus:2018mep}), and is actually irrelevant as it is  always  projected out in the final results. 
Using the  new connection~(\ref{Gammat2})   we can conveniently separate the $B$-field contributions and eventually  acquire the results~(\ref{SEPARATE}).

Now,   remembering   $\mP^{\mu\nu}=-\mbrP^{\mu\nu}=\half H^{\mu\nu}$~(\ref{circled}) and $\mP_{A}{}^{B}+\mbrP_{A}{}^{B}=\delta_{A}{}^{B}$,  we subtract the  two quantities in (\ref{cBcovD}),  and acquire   a  desired   covariant derivative, or  $\,\hfD^{\mu}=H^{\mu\rho}\partial_{\rho}+\hOmega^{\mu}$~(\ref{hOmega}):
\be
2\left[(\cB^{-1}P)^{\lambda C}(\cB^{-1}\brP)^{BD}-(\cB^{-1}\brP)^{\lambda C}(\cB^{-1}P)^{BD}\right]\na_{C}V_{D}=\left(\ba{l}
\hfD^{\lambda}\mV_{\mu}-\hPhi^{\lambda}{}_{\rho\mu}\mV^{\rho}\\
\hfD^{\lambda}\mV^{\nu}+\half\hH^{\lambda\nu\sigma}\mV_{\sigma}
\ea
\right),
\label{subtract}
\ee
where, with shorthand notation, 
\be
\ba{ll}
(\mP\hGamma\mbrP)_{ABC}:=\mP_{A}{}^{D}\hGamma_{DBE}\mbrP^{E}{}_{C}\,,
\qquad&\qquad
(\mbrP\hGamma\mP)_{ABC}:=\mbrP_{A}{}^{D}\hGamma_{DBE}\mP^{E}{}_{C}\,,
\ea
\ee
we set, extending (\ref{MhOhH}), 
\be
\ba{ll}
\hPhi^{\lambda}{}_{\mu\nu}=
2(\mP\hGamma\mbrP)^{\lambda}{}_{\mu\nu}-
2(\mbrP\hGamma\mP)^{\lambda}{}_{\mu\nu}\,,\qquad&\qquad
\deltaM \hPhi^{\lambda}{}_{\mu\nu}=-\deltaM B_{\mu\rho}\Omega^{\lambda\rho}{}_{\nu}+
\deltaM B_{\nu\rho}\hOmega^{\lambda\rho}{}_{\mu}\,,
\\
\hOmega^{\mu\nu}{}_{\lambda}=2(\mP\hGamma\mbrP)^{\mu\nu}{}_{\lambda}-
2(\mbrP\hGamma\mP)^{\mu\nu}{}_{\lambda}\,,\qquad&\qquad
\deltaM\hOmega^{\mu\nu}{}_{\lambda}=-\half\deltaM B_{\lambda\rho}\hH^{\mu\nu\rho}\,,\\
\hH^{\lambda\mu\nu}=4(\mP\hGamma\mbrP)^{\lambda\mu\nu}-
4(\mbrP\hGamma\mP)^{\lambda\mu\nu}\,,\qquad&\qquad \deltaM\hH^{\lambda\mu\nu}=0\,.
\ea
\label{OmegahH}
\ee
With $\partial_{A}=(0,\partial_{\mu})$ and  $\xi^{A}=(0,\xi^{\mu})$~(\ref{oLie}),    using Eq.(2.43) of \cite{Angus:2018mep},  we get under diffeomorphisms,
\be
\ba{l}
\delta_{\xi}(\mP\hGamma\mbrP)_{ABC}=\cL_{\xi}(\mP\hGamma\mbrP)_{ABC}
+\mP_{A}{}^{\rho}\mbrP_{C}{}^{\sigma}\partial_{\rho}\partial_{\sigma}\xi_{B}
-\mP_{A}{}^{\rho}\mbrP_{C\sigma}\partial_{\rho}\partial_{B}\xi^{\sigma}\,,\\
\delta_{\xi}(\mbrP\hGamma\mP)_{ABC}=\cL_{\xi}(\mbrP\hGamma\mP)_{ABC}
+\mbrP_{A}{}^{\rho}\mP_{C}{}^{\sigma}\partial_{\rho}\partial_{\sigma}\xi_{B}
-\mbrP_{A}{}^{\rho}\mP_{C\sigma}\partial_{\rho}\partial_{B}\xi^{\sigma}\,.
\ea
\ee
Hence both $\hPhi^{\lambda}{}_{\mu\nu}$ and $\hH^{\lambda\mu\nu}$ are  diffeomorphism covariant (and surely $\GL(n)\times \GL(\brn)$ invariant) tensors.

Further, due to  identities, 
\be
\ba{ll}
(\mP\hGamma\mbrP)_{A}{}^{\mu}{}_{B}=(\mbrP\hGamma\mP)_{B}{}^{\mu}{}_{A}\,,\qquad&\qquad
(\mP\hGamma\mbrP)^{\mu}{}_{(AB)}=(\mbrP\hGamma\mP){}^{\mu}{}_{(AB)}\,,
\ea
\ee
$\hH^{\lambda\mu\nu}$ and $\hPhi^{\lambda}{}_{\mu\nu}$ are skew-symmetric, 
\be
\ba{ll}
\hH^{\lambda\mu\nu}=\hH^{[\lambda\mu\nu]}\,,\qquad&\qquad\hPhi^{\lambda}{}_{\mu\nu}=
\hPhi^{\lambda}{}_{[\mu\nu]}\,,
\ea
\ee 
and we may express $\hOmega^{\mu\nu}{}_{\lambda}$  in different  ways,   
\be
\hOmega^{\mu\nu}{}_{\lambda}=-2(\mP\hGamma\mbrP)^{\mu}{}_{\lambda}{}^{\nu}+
2(\mbrP\hGamma\mP)^{\mu}{}_{\lambda}{}^{\nu}=
-2(\mP\hGamma\mbrP)_{\lambda}{}^{\nu\mu}+
2(\mbrP\hGamma\mP)_{\lambda}{}^{\nu\mu}\,.
\ee

In particular, when the circled vector, $\mV_{A}=(0,\mV_{\mu})$, is   derivative-index-valued  as ${\mV^{\mu}\equiv0}$, from (\ref{Mcov2}) $\mV_{\mu}$ becomes Milne-shift invariant and so does  $\hfD^{\lambda}\mV_{\mu\,}$,
\be
\ba{ll}
\deltaM\mV_{\mu}=0\,,\qquad&\qquad
\deltaM(\hfD^{\lambda}\mV_{\mu})=0\,.
\ea
\label{MSinv}
\ee

Alternative combination of (\ref{compCOV}), rather than (\ref{subtract}),  can give  different  type of covariant derivatives,
\be
\ba{ll}
(\fD_{i}V)^{\mu}:=
Y^{\rho}_{i}\!\left(H^{\mu\sigma}\partial_{\rho}V_{\sigma}-\Omega^{\mu\sigma}{}_{\rho}V_{\sigma}\right),\qquad&\qquad
(\brfD_{\bri}V)^{\mu}:=\brY^{\rho}_{\bri}\!\left(
H^{\mu\sigma}\partial_{\rho}V_{\sigma}-\Omega^{\mu\sigma}{}_{\rho}V_{\sigma}\right).
\ea
\label{YDT}
\ee
However, these can act only on one-form fields, and  appear  not so useful.

\newpage


\end{document}